\documentclass{aa}
\usepackage{graphicx}

\newcommand{\reff}{\mbox{$R_\mathrm{eff}$}}
\newcommand{\mfwhm}{\mbox{$\langle$FWHM$\rangle$}}
\newcommand{\malph}{\mbox{$\langle \alpha \rangle$}}
\newcommand{\msun}{\mbox{$M_{\odot}$}}
\newcommand{\rms}{\mbox{r.m.s.}}
\newcommand{\ssfr}{\mbox{$\Sigma_\mathrm{SFR}$}}
\newcommand{\deltav}{\mbox{$\langle \Delta V_\mathrm{Ground-WFPC2}\rangle$}}
\newcommand{\deltabv}{\mbox{$\langle \Delta (B\!-\!V)_\mathrm{Ground-WFPC2}\rangle$}}

\newcommand{\bv}{\mbox{$B\!-\!V$}}

\newcommand{\meanreff}{3.94}

\newcommand{\emeanreff}{0.12}

\newcommand{\slopefwhm}{0.08}
\newcommand{\eslopefwhm}{0.03}
\newcommand{\slopereff}{0.10}
\newcommand{\eslopereff}{0.03}

\newcommand{\constfwhm}{0.94}
\newcommand{\econstfwhm}{0.25}
\newcommand{\constreff}{1.12}
\newcommand{\econstreff}{0.35}

\begin{document}
\title{The Structure and Environment of Young Stellar Clusters in 
       Spiral Galaxies
  \thanks{Tables \ref{tab:all}, \ref{tab:more_comments}, \ref{tab:hstphot}, 
    and \ref{tab:gb} are only available in electronic form at the
    CDS via anonymous ftp to cdsarc.u-strasbg.fr (130.79.125.5) or
    via http://cdsweb.u-strasbg.fr/Abstract.html}\fnmsep
  \thanks{Based on observations obtained with the NASA/ESA
    \emph{Hubble Space Telescope}, obtained at the Space Telescope
	  Science Institute, which is operated by the Association
	  of Universities for Research in Astronomy, Inc., under
	  NASA contract NAS 5-26555}
}
\authorrunning{S. S. Larsen}
\titlerunning{Young Stellar Clusters in Spirals}

\author{S. S. Larsen}

\institute{European Southern Observatory (ESO),
  Karl-Schwarzschild-Str. 2, D-85748 Garching b. M{\"u}nchen, Germany,
  email: slarsen@eso.org}

\date{Received 17 Oct 2003; Accepted 03 Dec 2003}

\abstract{
A search for stellar clusters has been carried out in 18 nearby spiral 
galaxies, using archive images from the Wide Field Planetary Camera 2 on 
board the \emph{Hubble Space Telescope}.  All of the galaxies
have previously been imaged from the ground in $UBVI$. A catalogue of
structural parameters, photometry and comments based on visual inspection
of the clusters is compiled and used to investigate correlations between
cluster structure, environment, age and mass. 
Least-squares fits to the data suggest correlations between both
the full-width at half-maximum (FWHM) and half-light radius (\reff) of
the clusters and their masses ($M$) at about the $3\sigma$ level. Although 
both relations show a large scatter, the fits have substantially shallower 
slopes than for a constant-density relation (size $\propto$ $M^{1/3}$). 
However, many of the youngest clusters have extended halos which make the 
\reff\ determinations uncertain.
There is no evidence for galaxy-to-galaxy variations in the mean cluster 
sizes.  In particular, the mean sizes do not appear to depend on the host 
galaxy star formation rate surface density. 
Many of the youngest objects (age $< 10^7$ years) are located in strongly
crowded regions, and about 1/3-1/2 of them are double or multiple sources.
The HST images are also used to check the nature of cluster candidates 
identified in a previous ground-based survey. The contamination rate in
the ground-based sample is generally less than about 20\%, but some 
cluster identifications remain ambiguous because of crowding even with 
HST imaging, especially for the youngest objects.

}

\maketitle

\section{Introduction}

  In previous papers (Larsen \& Richtler \cite{lr99}; Larsen \cite{lar99};
Larsen \& Richtler \cite{lr00}; hereafter Paper I, II and III), we have
studied populations of young stellar clusters in the disks of nearby
spiral galaxies using ground-based imaging. That work was motivated by
a desire to understand why some galaxies host young stellar clusters
which are significantly more luminous (and, presumably, more massive)
than open clusters in the Milky Way. Well-known examples of galaxies
with rich populations of luminous young clusters include a number
of merger galaxies and starbursts (see e.g.\ compilation in 
Whitmore \cite{whit03}), but there are also some relatively ``normal'' 
galaxies such as the Large Magellanic Cloud and M33 which host a number 
of unusually (by Milky Way standards, at least) bright and massive young
clusters (Shapley \& Nail \cite{sn51}; Hodge \cite{hod61}; Richtler 
\cite{richtler93}; Christian \& 
Schommer \cite{cs82,cs88}; Chandar et al.\ \cite{chan99}). In Paper III 
we concluded that the main driving 
factor behind these differences seems to be the \emph{star formation rate} 
(SFR) of the host galaxy. Galaxies with high SFRs (per unit disk area) 
apparently form a larger fraction of their stars in massive, bound clusters.
The presence of highly luminous clusters in galaxies with high SFRs may
be --- at least partially --- a size-of-sample effect, due to the rich
cluster populations in such galaxies (Billett et al.\ \cite{bhe02}; 
Larsen \cite{lar02}).

  The study of stellar clusters is intimately linked to that of
star formation in general.
  Observations show that a large fraction, if not the vast majority, of 
all stars are born in clusters (e.g.\ Carpenter \cite{carp00}; 
Lada \& Lada \cite{lada03}). This does not, however, imply that all embedded
clusters are dynamically bound entities which survive emergence from 
their native molecular cloud cores and become observable at optical 
wavelengths. Lada \& Lada (\cite{lada03}) estimated that less
than 4--7\% of local embedded clusters survive to become bound
clusters of Pleiades age ($\sim10^8$ years), but this number
may depend on environment. In Paper III we noted a steady increase with 
host galaxy area-normalised star formation rate (\ssfr ) in the fraction 
of $U$-band light originating from clusters, ranging from well below 1\% in 
galaxies with very low SFRs (like IC~1613) to several percent in starbursts. 
Meurer et al.\ (\cite{meu95}) found that on average about 20\% of the UV 
light in a sample of starburst galaxies comes from young clusters. 
De Grijs et al.\ (\cite{deg03}) estimated that as much as $\sim70$\% 
of the B--band light in the tidal tails of the ``Tadpole'' and ``Mice'' 
galaxies originates from young clusters or compact star-forming regions.
In most ``normal'' star-forming galaxies, the young clusters contribute about 
1\% of the $U$-band light. Whether or not an embedded cluster remains
gravitationally bound depends on the star formation efficiency within
the proto-cluster cloud, as well as the timescale on which the gas is 
expelled (Elmegreen \cite{elm83}; Kroupa \cite{kroupa01}). It may therefore 
be more appropriate to 
view the fraction of optically visible young stars associated with clusters 
as a \emph{survival} frequency, associated with the star formation 
efficiency, than a cluster \emph{formation} efficiency \emph{per se} (which 
is probably always close to 100\%). 

  While care must be taken when interpreting the above results, due to 
possible differences in the age distributions of the clusters and/or field 
stars, completeness limits, etc., it seems clear that the Solar 
neighbourhood samples only a small part of the conditions under which 
star formation takes place in the Universe. Fortunately, there are
several star-forming galaxies available within a few Mpc, spanning a
range in SFRs, morphological type etc., which can be studied in
considerable detail with a combination of ground-based and space-based
techniques.  
The original sample of 21 nearby spirals analysed in Papers I--III 
has since been augmented by an additional handful of 
galaxies observed with the 3-m Shane telescope at Lick Observatory 
(see Larsen \cite{lar02}).  However, on the ground-based images, clusters 
were only marginally resolved, and although significant efforts were made to 
weed out the most obvious contaminants, the cluster lists in Paper II
should only be taken as provisionary. Many of the galaxies have 
now been imaged with the \emph{Wide Field Planetary Camera 2} (WFPC2) on board 
the \emph{Hubble Space Telescope} (HST) for a variety of reasons, often
with multiple pointings. The WFPC2 imaging not only provides a welcome
check of the true nature of the sources identified as cluster candidates
from the ground, but also allows the structure and immediate environment
of individual clusters to be examined in much greater detail.

  Several studies have indicated a remarkable uniformity in the sizes 
of stellar clusters over a wide range of masses, environments, ages and 
metallicities.  The most robust measure of cluster size is the
half-light or ``effective'' radius (\reff ) which is expected to remain
relatively stable over the lifetime of a cluster (Spitzer \cite{spit87}).
For Galactic globular clusters, \reff\ and luminosity are uncorrelated, 
although there is a trend of increasing cluster size with galactocentric 
distance ($R_{\rm GC}$; van den Bergh et al.\ \cite{van91}).  Using data 
from the compilation by Harris (\cite{har96}), the median \reff\ is 3.0 pc. 
Similarly, the diameters and masses of Galactic open clusters show no 
strong correlation, with typical sizes only slightly smaller
than those of globular clusters (Janes et al.\ \cite{jan88}). For young
clusters in the ``Antennae'' merger, Whitmore et al.\ (\cite{whit99}) found 
mean effective radii of $4\pm1$ pc. Zepf et al.\ (\cite{zepf99}) estimated
half-light radii of 5--10 pc for clusters in NGC~3256, perhaps slightly 
larger than for the Antennae, but again without any strong size-luminosity
correlation. For globular clusters around early-type galaxies, typical
effective radii are again 3--4 pc with no clear size-mass correlation
(e.g.\ Kundu \& Whitmore \cite{kw01}).  The lack of a significant mass-size 
relation is puzzling, since one might \emph{a priori} expect 
a cluster to form once the parent gas cloud reaches a certain density, 
independent of the total mass. If this initial gas density is reflected in 
the stellar density of the resulting cluster, one might naively expect the 
radius to scale with mass ($M$) roughly as $\reff\propto M^{1/3}$.  However, 
this is 
not what has generally been reported.  From the above examples, it appears 
that star clusters typically have effective radii of a few ($\sim3$) pc, with 
a scatter of perhaps 1--2 pc. Exceptions are found, however, including the 
faint ``Palomar''-type globular clusters in the outer part of the Galactic 
halo, and the ``faint fuzzy'' clusters recently discovered in a couple 
of nearby S0-type galaxies (Larsen \& Brodie \cite{lb00}; 
Brodie \& Larsen \cite{bl02}), which have larger effective radii ($\ga10$ pc).

  In this paper, the cluster candidates identified from the ground are first 
re-examined on archive WFPC2 images. Additional cluster candidates are then 
identified on the WFPC2 images and combined with ground-based photometry 
to produce a catalog of structural parameters and photometry for a sample 
of clusters. 
Relying on ground-based photometry limits the sample to relatively
bright objects, but has the advantage of providing uniform photometric
coverage of all clusters (even if crowding effects are more severe than
in the HST data). In particular, most of the HST datasets do not
include imaging in a $U$-band equivalent filter, which is essential for
age-dating the clusters. However, the HST photometry may still be
useful for some purposes and aperture photometry in an $r=0\farcs5$
aperture is presented for the available bandpasses in a separate table 
for each cluster candidate.
Each entry in the catalogue also contains 
various notes on the degree of crowding, close neighbours etc.\ based on a 
visual inspection of the candidates.  

  Because the HST images cover limited sections of the galaxies and span a 
huge range in exposure times and filters, the cluster sample presented here 
still cannot be considered complete in any sense. The completeness is
a complicated function of crowding, cluster size, underlying surface 
brightness, exposure time in the HST images, bandpass, seeing in the 
ground-based data, galaxy distance, and probably many other factors which
would be next to impossible to model in a satisfactory way.  It should also 
be emphasized that what is presented is still a list of cluster 
\emph{candidates}, which might contain contaminants (e.g.\ background
galaxies). A definitive list of bona-fide clusters would require spectroscopic 
follow-up, but such an effort is beyond the scope of this paper and left for 
future studies.  With this in mind, it is hoped that the catalogue may still 
provide a useful basis for further studies. As an example, it is used in 
Section~\ref{sec:full_sample} to investigate trends with age and mass 
in the cluster sizes, degree of crowding, and shape parameters.

\section{Data and Reduction}

\begin{table}
\caption{Exposures. $\Delta(\alpha,\delta)$ indicates the offsets
  in right ascension and declination between the coordinate systems of 
  the WFPC2 frames and the ground-based data (tied to the USNO catalogue). 
  The offsets are given in sec/15 for right ascension and in
  arcseconds for declination (i.e.\ the $\cos\delta$ factor has not
  been applied to convert the offsets in $\alpha$ to true arcsecs).}
\scriptsize
\label{tab:exposures}
\begin{tabular}{lllll} \hline
Galaxy   & Prop. & $\Delta(\alpha,\delta)$    & Filter & Exp.\\ \hline
\object{NGC 628} 
         & 9042  & $-0\farcs06, -0\farcs01$ & F450W  & $2\times230$ s  \\
	 &       &                          & F814W  & $2\times230$ s  \\
	 & 8597  & $-0\farcs05, +0\farcs01$ & F606W  & $160+400$ s     \\
	 & 5446  & $-0\farcs02, +0\farcs67$ & F606W  & $2\times80$ s   \\
	 & 9676  & $-0\farcs07, +0\farcs37$ & F300W  & $2\times1000$ s \\
	 &       &                          & F606W  & $3\times700$ s  \\
\object{NGC 1156}
         & 9124  & $-0\farcs04, -1\farcs05$ & F300W  & $2\times500$ s  \\
	 &       &                          & F814W  & $2\times40$ s   \\
\object{NGC 1313} 
         & 8599  & $+0\farcs05, -0\farcs89$ & F814W  & $2\times300+50$ s \\
	 & 9042  & $+0\farcs13, +0\farcs20$ & F450W  & $2\times230$ s \\
	 &       &                          & F606W  & $2\times230$ s \\
	 & 5446  & $+0\farcs72, +9\farcs43$ & F606W  & $2\times80$ s  \\
	 & 8199  &           -              & F814W  & $2\times1300$ s \\
	 & 6341  &           -              & F439W  & 60 s            \\
	 &       &                          & F555W  & 60 s            \\
	 & 6713  &           -              & F606W  & $2\times300$ s  \\
	 & 6802  &           -              & F814W  & $700+800+499$ s \\
\object{NGC 2835}
         & 9042  & $-0\farcs06, -0\farcs43$ & F450W  & $2\times230$ s   \\
	 &       &                          & F814W  & $2\times230$ s   \\
	 & 5446  & $-1\farcs10, -0\farcs50$ & F606W  & $2\times80$ s    \\
\object{NGC 2997}
         & 5446  & $-0\farcs07, -4\farcs03$ & F606W  & $2\times80$ s    \\
	 & 9042  & $-0\farcs06, +0\farcs19$ & F450W  & $2\times230$ s   \\
	 &       &                          & F814W  & $2\times230$ s   \\
\object{NGC 3184}
         & 5446  & $-0\farcs71, -3\farcs02$ & F606W  & $2\times80$ s    \\
	 & 8602  & $-0\farcs10, +0\farcs31$ & F555W  & $2\times350$ s   \\
	 & 9041  & $-0\farcs07, -1\farcs03$ & F555W  & $2\times80$ s    \\
	 &       &                          & F606W  & $4\times80$ s    \\
	 &       &                          & F675W  & $2\times80$ s    \\
	 &       &                          & F439W  & $2\times300$ s   \\
	 &       &                          & F814W  & $2\times300$ s   \\
\object{NGC 3521}
         & 5446  &             -            & F606W  & $2\times80$ s    \\
	 & 9042  &             -            & F450W  & $2\times230$ s   \\
	 &       &             -            & F814W  & $2\times230$ s   \\
	 & 5972  &             -            & F555W  & $2\times1000$ s  \\
\object{NGC 3621}
         & 5446  & $-1\farcs18, -5\farcs49$ & F606W  & $2\times80$ s    \\
	 & 5397  & $-0\farcs06, +1\farcs17$ & F555W  & $3\times900$ s   \\
	 &       &                          & F814W  & $4\times900$ s   \\
	 &       &                          & F439W  & $2\times900+180$ s \\
	 &       &                          & F336W  & $2100+1500+180$ s \\
         & 8584  & $+0\farcs08, +1\farcs00$ & F555W  & $4\times1300$ s  \\
	 &       &                          & F814W  & $4\times1300$ s  \\
\object{NGC 4258}
         & 8597  & $-0\farcs06, +0\farcs25$ & F606W  & $160+400$ s      \\
	 & 8591  & $-0\farcs04, -0\farcs49$ & F547M  & $4\times400$ s   \\
	 & 6888  & $-0\farcs07, +0\farcs28$ & F300W  & $2\times900+600$ s \\
	 & 5123  & $-0\farcs03, +0\farcs14$ & F547M  & $1000+160$ s      \\
	 & 9086  &             -            & F606W  & $3\times1300$ s   \\
	 &       &                          & F814W  & $3\times1300$ s   \\
	 & 8805  &             -            & F606W  & 1000 s            \\
	 & 7277  & $+0\farcs07, -0\farcs34$ & F555W  & $6\times500$ s    \\
	 &       &                          & F814W  & $6\times500$ s    \\
\object{NGC 5055}
         & 9042  & $-0\farcs11, +0\farcs15$ & F450W  & $2\times230$ s    \\
         &       &                          & F814W  & $2\times230$ s    \\
         & 5446  & $+0\farcs15, -1\farcs52$ & F606W  & $2\times80$ s     \\
	 & 8090  &             -            & F606W  & $500+700$ s       \\ 
\hline
\end{tabular}
\end{table}

\addtocounter{table}{-1}
\begin{table}
\caption{Exposures - continued}
\scriptsize
\begin{tabular}{lllll} \hline
Galaxy   & Prop.  & $\Delta(\alpha,\delta)$ & Filter & Exp. \\ \hline
\object{NGC 5194}
         & 5652   & $+0\farcs05, -0\farcs37$ & F336W  & $3\times400$ s    \\
	 & 5123   & $-0\farcs05, +0\farcs87$ & F547M  & $600+260$ s       \\
	 & 7375   & $+0\farcs02, +0\farcs05$ & F555W  & $2\times600$ s    \\
	 &        &                          & F814W  & $700+300$ s       \\
	 &        &                          & F336W  & $2\times600$ s    \\
	 &        &                          & F439W  & $600+500$ s       \\
         & 5777   & $-0\farcs13, -0\farcs54$ & F439W  & $2\times700$ s    \\
	 &        &                          & F555W  & 600 s             \\
	 &        &                          & F814W  & 600 s             \\
	 & 9073/1 & $-0\farcs06, -0\farcs46$ & F450W  & $4\times500$ s    \\
	 &        &                          & F555W  & $4\times500$ s    \\
	 &        &                          & F814W  & $4\times500$ s    \\
	 & 9073/2 & $+0\farcs00, +0\farcs55$ & F450W  & $4\times500$ s    \\
	 &        &                          & F555W  & $4\times500$ s    \\
	 &        &                          & F814W  & $4\times500$ s    \\
         & 5419   & $+0\farcs01, -0\farcs30$ & F547M  & 230 s             \\
	 & 9042   & $+0\farcs12, -0\farcs49$ & F606W  & $2\times230$ s    \\
	 &        &                          & F814W  & $2\times230$ s    \\
	 & 7909   &        -                 & F606W  & $2\times700$ s    \\
\object{NGC 5204}
         & 8601   & $+0\farcs10, +0\farcs23$ & F606W  & 600 s \\
	 &        &        -                 & F814W  & 600 s \\
\object{NGC 5236}
         & 5971   & 0        0               & F606W  & $1100+1200$ s     \\
	 &        &                          & F814W  & $2\times1000+2\times1300$ s \\
	 & 7909   & $+0\farcs07, -0\farcs02$ & F606W  & $2\times500$ s \\
	 & 8234   & $+0\farcs04, -0\farcs18$ & F547M  & $180+350+400$ s \\
	 &        &                          & F814W  & $160+200+350$ s \\
	 & 8805   & $+0\farcs03, +0\farcs48$ & F606W  & $2\times1000+2\times1400$ s \\
\object{NGC 5585}
         & 5446   & $+0\farcs69, -11\farcs08$ & F606W  & $2\times80$ s \\
	 & 8599   & $+0\farcs02, -0\farcs30$ & F814W  & $2\times300+40$ s \\
\object{NGC 6744}
         & 9042   & $-0\farcs17, +0\farcs69$ & F450W  & $2\times230$ s  \\
	 &        &                            & F814W  & $2\times230$ s  \\
	 & 8597   & $-0\farcs12, +1\farcs21$ & F606W  & $160+400$ s \\
	 & 5446   & $+4\farcs84, -20\farcs93$ & F606W  & $2\times80$ s \\
\object{NGC 6946}
         & 6118   & $+0\farcs10, -0\farcs05$ & F439W  & $2\times400$ s \\
	 &        &                          & F555W  & 400 s \\
	 & 8597   & $+0\farcs09, -0\farcs65$ & F606W  & $160+400$ s \\
	 & 8715   & $-0\farcs07, -0\farcs41$ & F439W  & $2\times1100$ s \\
	 &        &                          & F555W  & $2\times300$ s \\
	 &        &                          & F814W  & $2\times700$ s \\
\object{NGC 7424}
         & 9042   & $-0\farcs09, +0\farcs29$ & F450W  & $2\times230$ s \\
	 &        &                          & F814W  & $2\times230$ s \\
	 & 5446   & $-0\farcs67, -1\farcs47$ & F606W  & $2\times80$ s \\
	 & 8599   & $-0\farcs04, -1\farcs09$ & F814W  & $2\times300+40$ s \\
\object{NGC 7793}
         & 9042   & $-0\farcs10, -0\farcs02$ & F450W  & $2\times230$ s \\
	 &        &                          & F814W  & $2\times230$ s \\
         & 8591   & $-0\farcs08, -0\farcs28$ & F547M  & $4\times400$ s \\
	 & 8599   & $-0\farcs11, -0\farcs73$ & F814W  & $2\times300+40$ s \\
	 & 5446   & $-0\farcs23, -13\farcs25$ & F606W  & $2\times80$ s \\
	 & 8601   &        -                 & F606W  & 600 s \\
	 &        &                          & F814W  & 600 s \\ \hline
\end{tabular}
\end{table}

  The search for HST archive data was concluded in October 2002 and only 
includes WFPC2 datasets which had been publicly released up until that time
(Table~\ref{tab:exposures}).  ACS data were available for a few galaxies, 
but have been excluded in order to allow a relatively simple and 
homogeneous reduction procedure. 

  Given the large volume of data, a fairly high degree of automatization had 
to be incorporated in the reduction procedures. When several exposures
were available for a given field and filter, these were combined with
the CRREJ task in the STSDAS package in IRAF\footnote{IRAF is distributed 
by the National Optical Astronomical Observatories, which are operated by 
the Association of Universities for Research in Astronomy, Inc.~under 
contract with the National Science Foundation}. In most cases, no
shifts in the image coordinate systems were required before combination
but when necessary, such shifts were applied using the IMSHIFT command 
in IRAF. For each WFPC2 pointing and each filter, objects were then
detected with the DAOFIND task in DAOPHOT (Stetson \cite{stet87}) running 
within IRAF. Objects
with saturated pixels within a radius of 5 pixels were rejected.  The 
detected sources were fitted with
the ISHAPE profile-fitting algorithm (Paper II). ISHAPE models
each source assuming an analytic model for the intrinsic profile of
the source, convolved with the HST/WFPC2 point spread function (PSF).
The FWHM of the analytic model is iteratively adjusted until the best
fit to the data is obtained.  The initial round of profile fitting was
carried out with a fitting radius of $r=5$ pixels and assuming 
a model of the form
\begin{equation}
  P(r) \propto \left(1+(r/r_c)^2\right)^{-\alpha}
  \label{eq:moffat}
\end{equation}
with $\alpha=1.5$.  Profiles of this type were shown by Elson et al.\ 
(\cite{els87}) to provide good fits to young LMC clusters, and are 
hereafter referred to as ``EFF'' models (but note that the exponent 
$\alpha$ in Eq.~(\ref{eq:moffat}) corresponds to $\gamma/2$ in the 
definition by Elson et al.\ \cite{els87}).  The WFPC2 PSF was modelled 
using version 5 of the TINYTIM package (Krist \& Hook \cite{kh97}), including 
a convolution with the ``diffusion kernel'' to account for pixel-to-pixel 
charge diffusion.  PSFs were generated for each filter, and automatically 
selected by the reduction scripts to match the bandpasses used for the 
observations.

  Because the main aim was to study the structure of spatially 
resolved objects, only objects which were detected at $>10\sigma$ above
the background noise were included in the analysis.  Fainter objects would 
generally have too low S/N for reliable size measurements, and would have 
increased the already substantial computing time required to fit the spatial 
profiles. Because of the vastly different exposure times, different
bandpasses, background levels etc., the 10-$\sigma$ detection threshold
does not translate into a well-defined completeness limit in terms of
magnitude.  A total of 82000 sources were detected and fitted, requiring 
a few days of CPU time on a 1.5 GHz Pentium PC.  At this stage, many 
objects appeared several times in the source list, being included in 
multiple HST pointings and/or bands. 

  After the initial round of profile fitting, the HST object lists were 
matched with the photometry data files from the ground-based surveys (all 
details concerning the reduction of the ground-based data are given in Paper 
II).  Not only cluster candidates previously identified as such in the 
ground-based surveys, but \emph{all} point-like sources in each ground-based 
CCD frame for which photometry was available, were matched.  The matching was 
done by converting pixel coordinates measured on the WFPC2 images to 
celestial coordinates, using the METRIC task in the STSDAS package in IRAF. 
These coordinates were then matched with the coordinates of objects measured 
on the ground-based CCD images, tied to the US Naval Observatory meridian 
circle catalogue as provided by the ESO SkyCat Tool (Monet et al.\ 
\cite{mon98}). In many cases, there were clear systematic offsets between 
the WFPC2 and USNO coordinate systems. These offsets were determined by
displaying each individual WFPC2 frame and then marking the ground-selected 
cluster candidates contained within that frame. Because these would 
typically be among the brightest objects in the WFPC2 frames, and the 
offsets in general relatively minor ($\la 1\arcsec$), identifying the 
cluster candidates in the WFPC2 frames was usually unproblematic. The offsets 
between the WFPC2 and USNO coordinate systems are listed in 
Table~\ref{tab:exposures} for each dataset. Note that large offsets 
were found for the exposures belonging to snapshot programme 5446. 
This is probably because these datasets were guided using the ``Gyro Hold''
mode, which provides less accurate pointing and tracking than the 
Fine Guidance Sensors on HST.  A few WFPC2 pointings had no ground-based 
cluster candidates and were excluded from further analysis. For those
pointings, no offsets are given in Table~\ref{tab:exposures}.

\begin{table}
\caption{Explanation of comment codes}
\label{tab:comm_codes}
\begin{tabular}{ll} \hline 
a & double/multiple object of comparable brightness \\
b & fainter neighbours within $0\farcs5$ \\ 
c & companions of similar brightness between $0\farcs5<r<1\farcs5$ \\ 
d & elongated \\
e & chain of compact sources \\
f & irregular \\
g & crowded \\
h & in group/association of more than 5 bright stars/clusters \\
i & irregular background \\
j & saturated star \\
k & near edge of CCD \\
l & star? \\
m & galaxy? \\ \hline
\end{tabular}
\end{table}

\begin{figure}
\includegraphics[width=85mm]{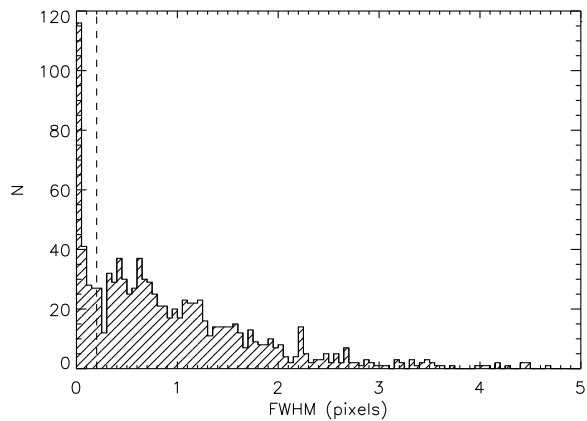}
\caption{Histogram of FWHM values for objects fitted during the first
 round of profile fitting. The narrow sequence of unresolved objects
 with FWHM$\approx$0 is clearly seen. The vertical dashed line
 indicates the criteria used to select cluster candidates.}
\label{fig:sz_sel}
\end{figure}

\begin{figure}
a (double/multiple): \\
n1313-540, n4258-1132, n5194-477 \\
\includegraphics[width=17mm]{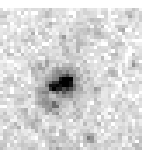}
\includegraphics[width=17mm]{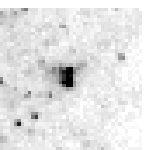}
\includegraphics[width=17mm]{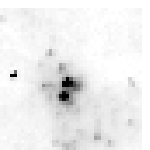} \\
b (fainter neighbours):\\
n1313-304, n4258-1051, n5194-239 \\
\includegraphics[width=17mm]{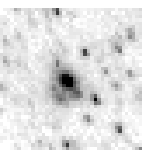}
\includegraphics[width=17mm]{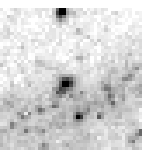}
\includegraphics[width=17mm]{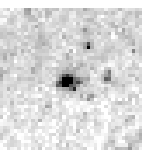} \\
c (companions within $1\farcs5$):\\
n1313-471, n4258-454, n5194-121 \\
\includegraphics[width=17mm]{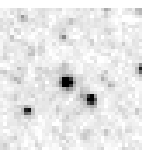}
\includegraphics[width=17mm]{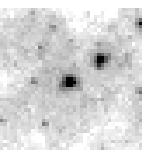}
\includegraphics[width=17mm]{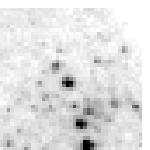} \\
d (elongated):\\
n2835-625, n3184-277, n5194-351 \\
\includegraphics[width=17mm]{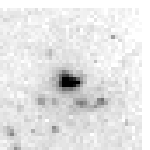}
\includegraphics[width=17mm]{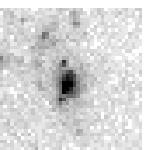}
\includegraphics[width=17mm]{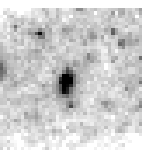} \\
e (chain):\\
n5194-424, n5194-906, n628-1895 \\
\includegraphics[width=17mm]{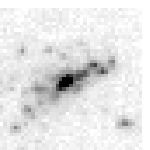}
\includegraphics[width=17mm]{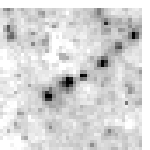}
\includegraphics[width=17mm]{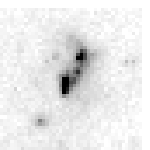} \\
f (irregular):\\
n1313-279, n2835-890, n7424-142 \\
\includegraphics[width=17mm]{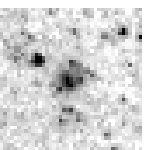}
\includegraphics[width=17mm]{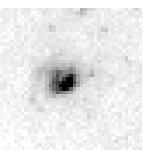}
\includegraphics[width=17mm]{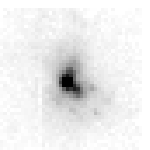} \\
g (crowded):\\
n5194-897, n1313-508, n2997-558 \\
\includegraphics[width=17mm]{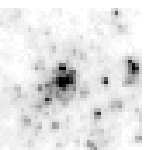}
\includegraphics[width=17mm]{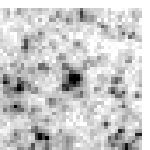}
\includegraphics[width=17mm]{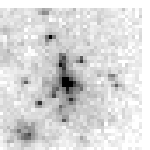} \\
h (in group of more than 5 stars/clusters):\\
n4258-776, n5194-433, n6946-1489 \\
\includegraphics[width=17mm]{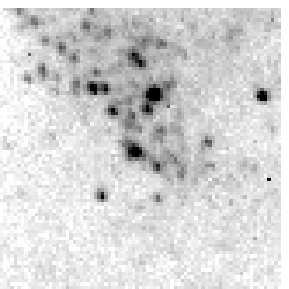}
\includegraphics[width=17mm]{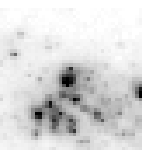}
\includegraphics[width=17mm]{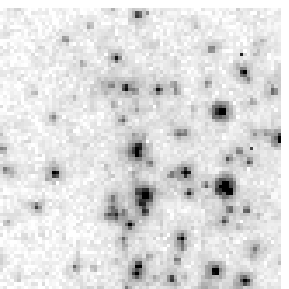} \\
i (irregular background):\\
n4258-1060, n4258-794, n5194-394 \\
\includegraphics[width=17mm]{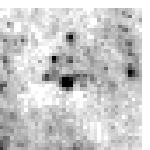}
\includegraphics[width=17mm]{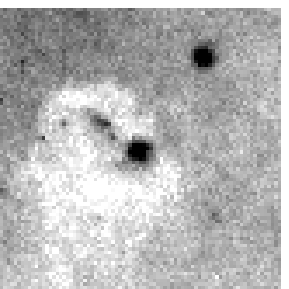}
\includegraphics[width=17mm]{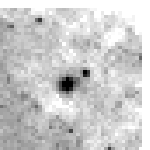}
\caption{Illustrations of comment codes a--i. Please see 
  Table~\ref{tab:comm_codes} for the full comments associated with
  each code. Each panel shows a $4\arcsec\times4\arcsec$ section
  around the cluster candidate.
  }
\label{fig:comm_ill}
\end{figure}

\begin{figure}
\includegraphics[width=20mm]{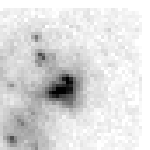}
\includegraphics[width=20mm]{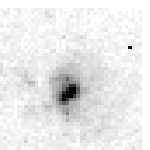}
\includegraphics[width=20mm]{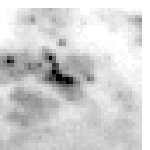}
\includegraphics[width=20mm]{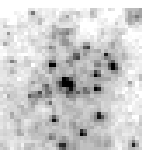}
\caption{Examples of objects classified as Type 2 (uncertain)}
\label{fig:uncertain}
\end{figure}

\begin{figure}
N3621--513:\\
\begin{minipage}{20mm}
  F336W\\
  \includegraphics[width=20mm]{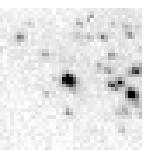}
\end{minipage}
\begin{minipage}{20mm}
  F439W \\
  \includegraphics[width=20mm]{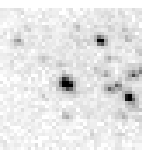}
\end{minipage}
\begin{minipage}{20mm}
 F555W
 \includegraphics[width=20mm]{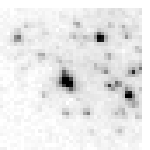}
\end{minipage}
\begin{minipage}{20mm}
F814W\\
\includegraphics[width=20mm]{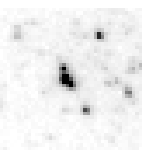}
\end{minipage} \\
N3621-620:\\
\begin{minipage}{20mm}
  F336W\\
\includegraphics[width=20mm]{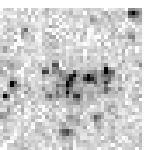}
\end{minipage}
\begin{minipage}{20mm}
  F439W\\
\includegraphics[width=20mm]{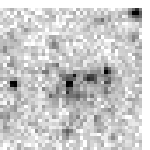}
\end{minipage}
\begin{minipage}{20mm}
  F555W\\
\includegraphics[width=20mm]{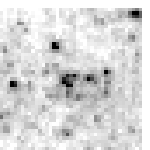}
\end{minipage}
\begin{minipage}{20mm}
  F814W\\
\includegraphics[width=20mm]{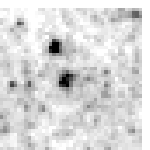}
\end{minipage} \\
\caption{Two examples of how the appearance of cluster candidates can 
  change dramatically depending on bandpass.}
\label{fig:notes_objects}
\end{figure}

  The initial round of profile fitting served to separate point-like sources 
(likely stars) from extended ones (cluster candidates).  Fig.~\ref{fig:sz_sel} 
shows the distribution of intrinsic FWHM values (in pixels). There is
a very narrow peak of objects with FWHM values close to 0, although there 
is no sharp division between resolved and unresolved objects.  For further 
analysis, a total of 3100 sources with S/N$>$50 (measured within the 5 
pixels aperture on the WFPC2 images), FWHM$>$0.2 pixels (dashed line in 
Fig.~\ref{fig:sz_sel}) and ground-based photometry were selected. For 
comparison, the (undersampled) PSF of WFPC2 itself has a FWHM of about 1.5 
pixels.  At a typical distance of 5 Mpc, the size cut corresponds to a FWHM 
of 0.5 pc or a core radius of about 0.25 pc.  Many of these sources still 
represented multiple observations of the same object.  
Objects satisfying these criteria were again 
fitted, but this time allowing the power-law index ($\alpha$) of the EFF 
profiles to vary as a free parameter. In order to better constrain 
$\alpha$, a fitting radius of 10 instead of 5 pixels was 
chosen for this second round. The choice of a 10 pixels fitting radius 
represents a compromise between reasonably accurate constraints on the fit 
parameters, reducing the effects of a non-uniform background, and 
computing time.
Each source was fitted three times, using 
different initial guesses for $\alpha$ ($\alpha_0=1.0$, 1.5 and 2.0). 
The values 1.0 and 1.5 roughly bracket those typically found for real star
clusters (e.g.\ Elson et al.\ \cite{els87} and Section~\ref{sec:struct_env}), 
while an additional more extreme value of $\alpha_0=2$ was included in
order to reduce any \emph{a-priori} bias in the measurements, in
case some clusters have steeper slopes (although tests of the
ISHAPE algorithm suggest that any dependence of the fitted parameters
on the input guesses is relatively minor; see Sec \ref{sec:ishape_tests} 
below).

  Finally, all cluster candidates fitted in the second round
were visually inspected. In addition, \emph{all} cluster candidates
identified in the ground-based surveys and with HST imaging were
inspected regardless of the S/N in the HST images. The inspection was
done by displaying a $4\arcsec\times4\arcsec$ section of all images of 
each cluster candidate in an IDL widget, which also
contained a number of check-boxes corresponding to the comment
codes given in Table~\ref{tab:comm_codes}. Each
object was also assigned to one of three types: 1) likely/certain
cluster, 2) uncertain classification, 3) likely/certain non-cluster.
A flag was set if the ISHAPE fit was unlikely to have resulted
in meaningful structural parameters, as judged from the degree of
crowding etc. The inspection process was repeated three times, each
time displaying the cluster candidates in a different (random)
order. At the end, the final object type and ISHAPE ``Fit-OK'' flag
were determined as the most pessimistic of the three individual
estimates. A comment code was included in the final list if
selected in at least 2 out of the 3 inspection rounds. 

  Clearly, by its very nature any visual inspection involves an element 
of subjectivity. Very often, more than one comment applied to
a given object. In such cases, all applicable comments were selected and
included in the catalogue.  Furthermore, resolution (and thus distance) 
effects may change the visual appearance of certain objects and
thereby e.g.\ cause an object classified as ``double'' in a nearby galaxy
to be labelled ``elongated'' in a slightly more distant one. In order
to illustrate roughly what the individual comment codes in
Table~\ref{tab:comm_codes} represent, Fig.~\ref{fig:comm_ill} shows
$4\arcsec\times4\arcsec$ sections around some clusters that are
typical for each comment code. Fig.~\ref{fig:uncertain} shows a few
examples of objects classified as ``uncertain''. Another difficulty 
is that objects can appear quite different in different bandpasses.
As an example, Fig.~\ref{fig:notes_objects} shows two
objects observed in F336W, F439W, F555W and F814W. The F336W and
F814W images of N3621-620 are hardly recognisable as the same
object --- the F336W image just shows what might be a loose
association of stars, while the F814W image shows two fairly well-defined,
compact sources. Conversely, N3621-513 looks compact and symmetric
in F336W, while two nearby neighbours (perhaps red giants or
supergiant stars) appear in F814W. If multi-wavelength data were available
for all objects, one might adopt a consistent strategy for dealing with 
this problem, but in many cases data were only available in one band.
It should therefore be borne in mind that the visual comment codes
(and even the profile fits) depend on the bandpass in 
a way that is not easily predictable.

\begin{table*}
  \caption{All clusters with S/N$>$50 on HST images.  Photometry is from 
  ground-based data while morphological data are from HST/WFPC2 images.  
  The first 10 rows of the table are reproduced here; the full table 
  (1358 rows) is only available in electronic form.}
\tiny
\label{tab:all}
\begin{tabular}{lrrrrrrrr} \hline
ID & R.A. (2000.0) & Decl. (2000.0) & $V$ & $U-B$ & $B-V$ & $V-I$ & $R_{\rm GC}$ & \\
   & $N$ & FWHM (pc) & $\alpha$ & R$_{\rm eff}$ (pc) & $x/y$ & Code & Fit ok? & Comments \\ 
(1)& (2)           & (3)            & (4) & (5)   & (6)   & (7)   & (8)          &  \\
   & (9)&  (10)     & (11)  &  (12)              & (13)  & (14) & (15)    & (16)  \\ \hline
n1156-486 & 2:59:41.29 &  25:14:15.34& $16.97\pm0.01$ & $-0.96\pm0.01$ & $ 0.44\pm0.01$ & $-0.53\pm0.02$ & $ 0.30\arcmin$ & \\
& 1 & $8.55\pm0.81$ & $0.95\pm0.08$ & $12.42\pm1.24$ & $0.63\pm0.04$ & 1 & N & a,f \\
n1156-583 & 2:59:42.85 &  25:14:28.68& $17.41\pm0.01$ & $-0.29\pm0.01$ & $ 0.29\pm0.01$ & $ 0.65\pm0.01$ & $ 0.23\arcmin$ & \\
& 1 & $2.03\pm0.55$ & $0.88\pm0.05$ & $10.59\pm1.11$ & $0.97\pm0.03$ & 1 & Y & - \\
n1313-180 & 3:18:15.12 &  $-$66:31:55.29& $20.67\pm0.04$ & $ 0.11\pm0.15$ & $ 0.67\pm0.07$ & $ 0.98\pm0.06$ & $ 2.07\arcmin$ & \\
& 1 & $5.35\pm0.11$ & $0.89\pm0.01$ & $13.57\pm0.53$ & $0.95\pm0.03$ & 1 & Y & - \\
n1313-199 & 3:18:17.49 &  $-$66:31:36.73& $20.31\pm0.03$ & $ 0.03\pm0.08$ & $ 0.18\pm0.04$ & $ 0.49\pm0.05$ & $ 1.77\arcmin$ & \\
& 1 & $10.12\pm0.68$ & $2.09\pm0.58$ & $6.92\pm0.59$ & $0.78\pm0.03$ & 1 & Y & d \\
n1313-228 & 3:18:18.46 &  $-$66:31:22.76& $19.30\pm0.02$ & $ 0.23\pm0.06$ & $ 0.42\pm0.03$ & $ 0.57\pm0.02$ & $ 1.56\arcmin$ & \\
& 1 & $1.15\pm0.47$ & $5.64\pm5.24$ & $0.52\pm0.16$ & $0.53\pm0.39$ & 1 & Y & - \\
n1313-234 & 3:18:07.67 &  $-$66:31:19.06& $17.87\pm0.01$ & $-0.82\pm0.01$ & $ 0.34\pm0.01$ & $-0.78\pm0.02$ & $ 1.66\arcmin$ & \\
& 1 & $9.24\pm2.37$ & $0.87\pm0.60$ & $19.43\pm6.90$ & $0.92\pm0.02$ & 2 & N & a,f,g \\
n1313-239 & 3:18:17.59 &  $-$66:31:12.41& $19.86\pm0.04$ & $-0.07\pm0.08$ & $ 0.26\pm0.06$ & $ 0.38\pm0.08$ & $ 1.37\arcmin$ & \\
& 4 & $6.34\pm0.46$ & $1.29\pm0.32$ & $8.90\pm1.09$ & $0.90\pm0.02$ & 1 & Y & - \\
n1313-248 & 3:18:16.05 &  $-$66:31:06.91& $18.85\pm0.03$ & $ 1.21\pm0.24$ & $ 1.03\pm0.07$ & $ 1.11\pm0.04$ & $ 1.27\arcmin$ & \\
& 1 & $2.14\pm0.22$ & $70.80\pm90.78$ & $0.67\pm0.05$ & $0.22\pm0.03$ & 2 & Y & - \\
n1313-249 & 3:18:12.45 &  $-$66:31:06.38& $19.95\pm0.06$ & $ 0.10\pm0.10$ & $ 0.20\pm0.07$ & $ 0.60\pm0.08$ & $ 1.29\arcmin$ & \\
& 3 & $5.18\pm0.26$ & $1.25\pm0.08$ & $7.09\pm0.70$ & $0.92\pm0.03$ & 1 & Y & - \\
n1313-259 & 3:18:15.55 &  $-$66:31:02.25& $20.71\pm0.14$ & $-0.31\pm0.19$ & $ 0.26\pm0.19$ & $ 0.71\pm0.18$ & $ 1.19\arcmin$ & \\
& 2 & $1.40\pm0.42$ & $1.41\pm0.20$ & $1.44\pm0.05$ & $0.53\pm0.03$ & 1 & Y & - \\
\hline
\end{tabular}
\end{table*}

\begin{table}
  \caption{Additional comments (only 5 sample entries given)}
\label{tab:more_comments}
\begin{tabular}{lp{60mm}}
ID  &  Comments \\ \hline
n1313-356  & identification uncertain \\
n1313-409  & tracking error \\
n1313-411  & cluster members resolved? \\
n2835-677  & very elongated halo \\
n2997-616  & bad pixels \\
\hline
\end{tabular}
\end{table}

\begin{table*}
  \caption{HST photometry for the same clusters listed in Table~\ref{tab:all}.
     All magnitudes are in the STMAG system, measured in a $0\farcs5$ 
     aperture and applying an aperture correction of $-0.2$ mag. No 
     corrections for foreground reddening have been applied.  
     Only the first 10 rows are reproduced here.}
\label{tab:hstphot}
\begin{tabular}{lcccccccc} \hline
ID & F300W & F336W & F439W & F450W & F547M & F555W & F606W & F814W \\ \hline
n1156-486  & $18.29\pm0.01$  &       ---       &       ---       &       ---       &       ---       &       ---       &       ---       &       ---      \\
n1156-583  & $19.68\pm0.04$  &       ---       &       ---       &       ---       &       ---       &       ---       &       ---       & $18.90\pm0.01$ \\
n1313-180  &       ---       &       ---       &       ---       &       ---       &       ---       &       ---       & $21.11\pm0.01$  &       ---      \\
n1313-199  &       ---       &       ---       &       ---       &       ---       &       ---       &       ---       & $21.05\pm0.01$  &       ---      \\
n1313-228  &       ---       &       ---       &       ---       & $19.51\pm0.01$  &       ---       &       ---       & $19.66\pm0.01$  & $20.18\pm0.01$ \\
n1313-234  &       ---       &       ---       &       ---       &       ---       &       ---       &       ---       & $19.28\pm0.01$  &       ---      \\
n1313-239  &       ---       &       ---       &       ---       & $20.38\pm0.02$  &       ---       &       ---       & $20.62\pm0.01$  & $21.01\pm0.01$ \\
n1313-248  &       ---       &       ---       &       ---       & $19.32\pm0.01$  &       ---       &       ---       & $18.86\pm0.00$  & $18.96\pm0.00$ \\
n1313-249  &       ---       &       ---       &       ---       & $20.28\pm0.02$  &       ---       &       ---       & $20.56\pm0.01$  & $20.98\pm0.01$ \\
n1313-259  &       ---       &       ---       &       ---       &       ---       &       ---       &       ---       & $21.09\pm0.02$  & $21.52\pm0.02$ \\
\hline
\end{tabular}
\end{table*}

  The final list of cluster candidates includes 1358 objects
(Table~\ref{tab:all}). Columns (1)--(8) contain information from the 
ground-based data: Coordinates, $UBVI$ photometry and projected 
galactocentric distance (in arc\-min). The photometry has been corrected 
for Galactic foreground extinction using the Schlegel et al.\ 
(\cite{sch98}) values and the extinction law in Cardelli et al.\ 
(\cite{car89}). Note that the reddening corrections differ from those in 
Paper I--III, where Burstein \& Heiles (\cite{bh84}) values were used.  
In Paper I--III we used a relatively large aperture radius of 8 pixels 
($1\farcs5$ and $3\farcs2$ for the NOT and Danish 1.54 m data, respectively) 
to avoid possible systematic effects in the integrated magnitudes due to the 
extendedness of the objects.  However, it is now clear that most clusters are 
compact enough that this is not a major source of concern in ground-based 
imaging, and in the present paper I therefore use a smaller aperture radius 
of 4 pixels, the same as for the colours, for the ground-based magnitudes.
Information derived from HST images is listed in columns (9)--(16): The 
number of individual detections of each cluster 
(where one ``detection'' is defined as the presence of the cluster in
 an image taken through a given filter under a given programme),
$N$, is in col.\ (9), 
followed by the full-width at half maximum (FWHM) of the cluster profile 
derived from the ISHAPE fits, the exponent $\alpha$, the effective 
(half-light) radius \reff , the minor/major axis ratio ($x/y$), the object 
type (Col.\ 14), Fit-OK flag (Col.\ 15) and comments. 

  Unlike the classical King profiles (King \cite{king62,king66}), the 
EFF models have no finite radius, and for $\alpha\leq1$ the volume 
contained under the profile is infinite. For $\alpha$ only slightly larger 
than unity, the total volume converges very slowly, resulting in 
unrealistically large \reff\ values. To cope with these difficulties, the 
\reff\ values in Table~\ref{tab:all} are computed for a finite outer radius
of 50 pc, beyond which the luminosity profiles of young clusters become 
difficult to trace even in nearby galaxies such as the LMC 
(e.g.\ Elson et al.\ \cite{els87}). However, it is important to note
that the estimates of \reff\ are generally based on an extrapolation of the
luminosity profiles beyond the fitting radius, and carry significant
uncertainties especially when $\alpha\la1$.
  Instead of listing the FWHM, I 
could have given the core radius $r_c$, since both are always defined. The 
reason for listing FWHM is that there is some ambiguity in the definition 
of the core radius. Some authors define it as FWHM/2, but it may also be 
defined e.g.\ as the scale radius $r_c$ in Eq.~(\ref{eq:moffat}).  
In order to avoid confusion, I will simply use the term FWHM rather than 
``core radius'' throughout the remainder of this paper.  When more than 
one exposure was available for a cluster candidate, the shape parameters 
in Table~\ref{tab:all} were obtained by weighting the measurements on 
each exposure by its S/N.
As discussed above, the morphology of clusters can be quite
wavelength-dependent, but cases where the determination of shape
parameters is particularly uncertain can generally be recognized by 
the 'Fit-OK' flag in Column (15) of Table~\ref{tab:all}.

Errors were estimated as follows: For each exposure, the error
on the shape parameters derived from that exposure were estimated
as the standard deviation of the three individual fits. If only
one exposure was available, this is the error listed in 
Table~\ref{tab:all}. When several exposures were available, the
errors in Table~\ref{tab:all} are the estimated standard
errors on the mean of the weighted average.  Some additional
comments for a few objects, which did not fit into the codes in
Table~\ref{tab:comm_codes}, are listed in Table~\ref{tab:more_comments}.

\subsection{HST versus ground-based photometry}
\label{sec:photcmp}

HST photometry for each cluster candidate is given in Table~\ref{tab:hstphot}. 
The HST magnitudes are given in the STMAG system, since transformation to the 
Johnson-Cousins system requires colour information which is not always 
available. The photometry was obtained with the PHOT task in DAOPHOT, using 
an $r=0\farcs5$ aperture for the photometry and an $r=1\farcs0-1\farcs5$
annulus for the background measurements. The $0\farcs5$ aperture
contains about 90\% of the light from a point source (e.g.\ Holtzman et al.\ 
\cite{hol95}), but because the objects measured here are extended, an even 
larger fraction of the total light will fall outside the $0\farcs5$ aperture. 
Thus, a total aperture correction of $-0.2$ mag has been applied to the 
magnitudes in Table~\ref{tab:hstphot}. The exact correction will depend on 
the detailed spatial profiles of the clusters. In principle, this dependence 
might be estimated from the ISHAPE profile fits, but tying the photometry in 
Table~\ref{tab:hstphot} to the size measurements would make it difficult to 
backtrack these corrections. A constant $-0.2$ mag aperture correction is 
probably not far from the truth (Larsen \cite{lar02}), and more sophisticated
corrections can easily be applied if desired.
In cases where several observations were available for a cluster in a given 
band, Table~\ref{tab:hstphot} lists the measurement with the smallest 
formal error.

  In previous papers it has been documented that no systematic differences 
seem to be present between ground-based and HST-based \emph{colours} for the 
cluster candidates (e.g.\ Larsen \cite{lar02}), although a random scatter of 
0.1--0.2 mag exists. For integrated magnitudes, on the other hand, an
offset of a few $\times0.1$ mag has been found between ground-based and HST 
magnitudes, in the sense that ground-based data tends to give brighter 
magnitudes. With the larger sample of clusters available here, this
comparison can now be carried out in more detail. Of the clusters listed
in Table~\ref{tab:hstphot}, 1245 have data in at least one of the
filters F547M, F555W and F606W, all of which are reasonable approximations
to the Johnson $V$-band. The mean difference \deltav\ between ground-based
and WFPC2 photometry, including all objects with F547M, F555W or F606W
HST data is $-0.50$ mag with a large scatter of $\sigma (\deltav) = 0.74$ mag.
This scatter is partly due to the fact that the ground-based
photometry of some of the fainter clusters has large errors, but 
decreases only slightly (to $\sigma(\deltav) = 0.65$ mag) if
clusters with formal errors larger than 0.2 mag on the ground-based $V$
magnitudes are excluded. Thus, most of the errors are clearly of a
systematic nature. 

  If the samples observed with the Danish 1.54 m telescope and the NOT are 
compared with the HST photometry separately, interesting differences emerge. 
The image quality of the data taken with the two telescopes differ 
significantly, with typical FWHM seeing values of $1\farcs5$ and $0\farcs8$, 
respectively (Larsen \cite{lar99}). For the galaxies observed with the
Danish 1.54 m, the mean difference between ground-based and HST
photometry is $\deltav = -0.76\pm0.61$ mag (where the 0.61 mag refer to
the standard deviation around the mean, not the error on the mean value).
For the NOT sample, the corresponding numbers are $\deltav = -0.29\pm0.50$ mag.
Thus, while the scatter remains large even in the NOT data, the systematic
difference relative to the HST photometry is clearly smaller than for the
Danish 1.54 m data.
A few individual, relatively isolated clusters observed with both the NOT 
and Danish 1.54 m telescope have been analysed in detail 
(Larsen et al.\ \cite{laretal01}; Larsen \& Richtler 2004, in prep.) and for 
these clusters there is good agreement between ground-based and HST 
magnitudes (within $\sim0.1$ mag). The differences between the
mean magnitudes of the HST and ground-based samples are therefore not due 
to trivial zero-point errors in the photometric calibrations.

  Clusters for which any of the comment flags in Table~\ref{tab:comm_codes}
are set might be expected to show poorer agreement with the ground-based
data. Indeed, if such clusters are rejected then $\deltav = -0.40\pm0.61$ mag
for all clusters, and $\deltav=-0.59\pm0.59$ mag and $\deltav=-0.10\pm0.42$
mag for the Danish 1.54 m and NOT samples, respectively. Thus, the systematic
difference between HST and ground-based photometry clearly decreases,
albeit still with significant scatter. The remaining offsets can probably be 
attributed to contamination within the ground-based apertures which did 
not trigger any comment flags. In fact, if the HST photometry is instead
carried out using an $r=1\farcs0$ aperture radius, assuming that such an
aperture encircles all flux from the objects (i.e.\ applying no
aperture corrections) then the mean offset with respect to the NOT data
is only $\deltav=-0.03\pm0.35$ mag. Excluding clusters with comment codes,
the difference decreases even further to $\deltav=-0.003\pm0.33$ mag.

  A smaller subset of the clusters have observations in HST bandpasses
that allow a comparison with the ground-based \emph{colours}.  For example,
Holtzman et al.\ (\cite{hol95}) give transformations to Johnson \bv\
colours for the F439W and F555W bandpasses, which are available for
190 of the clusters in Table~\ref{tab:hstphot}. The mean offset
between ground-based and HST \bv\ colours is $\deltabv=-0.003$ mag
with a scatter of 0.17 mag. For the Danish 1.54 m and NOT samples, the
differences are $\deltabv=0.05\pm0.15$ mag and $\deltabv=-0.05\pm0.15$ mag,
respectively. This confirms that the ground-based colours are more accurate 
than magnitudes, presumably because the objects that contaminate the 
ground-based apertures tend to have similar ages and colours to the 
clusters themselves. 




\subsection{Tests of the profile-fitting algorithm}
\label{sec:ishape_tests}

\begin{table*}
\caption{Tests of the ISHAPE profile-fitting algorithm. For each
  combination of input parameters ($V=20/21/22$, FWHM=1pc /2 pc)
  the output fitted FWHM and $\alpha$ parameters are 
  shown for three initial guesses of $\alpha$ ($\alpha_0 = 1.0/1.5/2.0$).
  Numbers in parantheses denote the object-to-object \rms\ deviation 
  around the mean values, excluding the two most deviating points at 
  each extreme.
}
\tiny
\label{tab:ishape_tests}
\begin{tabular}{llll|llllllll} \hline
\multicolumn{4}{c|}{Input parameters} & 
  \multicolumn{8}{c}{Output parameters} \\
 & & & 
   & \multicolumn{2}{c}{$\alpha_0$ = 1.0}
   & \multicolumn{2}{c}{$\alpha_0$ = 1.5}
   & \multicolumn{2}{c}{$\alpha_0$ = 2.0}
   & \multicolumn{2}{c}{$\langle \sigma_3 \rangle$} \\
$V$ & FWHM, pix (pc) & $\alpha$  & 
  $\langle$S/N$\rangle$ & \mfwhm\ & \malph\ & \mfwhm\ & \malph\ & \mfwhm\ & \malph\ & FWHM & $\alpha$ \\ \hline
\multicolumn{4}{l|}{NGC 7793 (p8591, F547M)} \\
20.0 & 0.62 (1.0) & 1.0 &  95.4 & 0.65 (0.10) & 1.01 (0.06) & 0.67 (0.10) & 1.01 (0.05) & 0.64 (0.10) & 1.00 (0.06) & 0.039 & 0.023 \\
20.0 & 1.25 (2.0) & 1.0 &  96.7 & 1.27 (0.14) & 0.99 (0.08) & 1.29 (0.15) & 1.00 (0.08) & 1.27 (0.13) & 0.99 (0.07) & 0.033 & 0.019 \\
21.0 & 0.62 (1.0) & 1.0 &  40.0 & 0.79 (0.15) & 1.07 (0.10) & 0.76 (0.16) & 1.06 (0.11) & 0.80 (0.21) & 1.08 (0.13) & 0.076 & 0.044 \\
21.0 & 1.25 (2.0) & 1.0 &  41.1 & 1.25 (0.25) & 1.01 (0.13) & 1.25 (0.27) & 1.00 (0.13) & 1.22 (0.31) & 0.99 (0.15) & 0.086 & 0.044 \\
22.0 & 0.62 (1.0) & 1.0 &  15.9 & 0.93 (0.29) & 1.23 (0.32) & 1.00 (0.28) & 1.36 (0.49) & 0.96 (0.33) & 1.24 (0.35) & 0.107 & 0.130 \\
22.0 & 1.25 (2.0) & 1.0 &  17.1 & 1.35 (0.48) & 1.15 (0.46) & 1.41 (0.44) & 1.17 (0.35) & 1.49 (0.45) & 1.17 (0.41) & 0.148 & 0.139 \\
\multicolumn{4}{l|}{NGC 5194 (p7375, F555W)} \\
20.0 & 0.25 (1.0) & 1.0 &  133.6 & 0.33 (0.06) & 1.03 (0.05) & 0.31 (0.05) & 1.03 (0.05) & 0.31 (0.06) & 1.02 (0.05) & 0.037 & 0.024 \\
20.0 & 0.50 (2.0) & 1.0 &  134.6 & 0.57 (0.09) & 1.02 (0.07) & 0.56 (0.10) & 1.02 (0.07) & 0.56 (0.07) & 1.02 (0.05) & 0.033 & 0.019 \\
21.0 & 0.25 (1.0) & 1.0 &  57.4 & 0.41 (0.17) & 1.11 (0.18) & 0.37 (0.12) & 1.08 (0.14) & 0.39 (0.12) & 1.07 (0.13) & 0.054 & 0.039 \\
21.0 & 0.50 (2.0) & 1.0 &  59.3 & 0.64 (0.13) & 1.04 (0.12) & 0.60 (0.13) & 1.04 (0.12) & 0.61 (0.11) & 1.05 (0.13) & 0.068 & 0.040 \\
22.0 & 0.25 (1.0) & 1.0 &  26.0 & 0.54 (0.19) & 1.21 (0.36) & 0.57 (0.24) & 1.38 (0.82) & 0.58 (0.20) & 1.39 (0.78) & 0.080 & 0.112 \\
22.0 & 0.50 (2.0) & 1.0 &  25.2 & 0.86 (0.20) & 1.24 (0.36) & 0.85 (0.20) & 1.23 (0.33) & 0.89 (0.27) & 1.29 (0.48) & 0.100 & 0.113 \\
\hline
\end{tabular}
\end{table*}

  Tests of the ISHAPE code have been carried out in several previous
papers, in particular in Paper II. The reader is
referred to that paper and to the documentation included with the
code for further details. However, because the ability to fit the 
EFF $\alpha$ parameter is a more recent addition, some tests of 
this particular feature are presented in the following.

  First, artificially generated sources with known profiles were added 
to images of NGC~7793 and NGC~5194. The frames used for these tests 
were the WF4 chips from proposal 8591 (NGC 7793, F547M, 4$\times$400 s) and 
proposal 7375 (NGC 5194, F555W, 2$\times$600 s).  The artificial sources 
were generated by convolving EFF profiles with the HST PSF, generated 
by TinyTim, and then added to the science images with the MKSYNTH program 
(Paper II). For each galaxy, artificial sources with FWHM of 1 and 2 pc, 
$\alpha$ = 1.0 and 1.5, and magnitudes of $V=20, 21$ and 22 were added. For 
each combination of these input parameters, 25 objects were added at random 
positions and then fitted with ISHAPE using three different values 
($\alpha_0$ = 1.0, 1.5 and 2.0) for the initial guess of $\alpha$. 

  The test results are summarised in Table~\ref{tab:ishape_tests}. For 
brevity, results are shown for $\alpha=1.0$ only, but the conclusions
reached for $\alpha=1.5$ are essentially identical. The first
four columns in each row give the input $V$ magnitude of the
test objects, the FWHM in pixels and pc, and the $\alpha$ parameter.
The remaining columns list the mean S/N of the test objects (within the 
fitting radius), followed by the fitted FWHM and $\alpha$ values for the 
three initial guesses, $\alpha_0$, of the $\alpha$ parameter. The numbers 
in parantheses denote the \rms\ deviation around the mean, excluding the 
two most deviating data points at each extreme. The last two columns list 
the mean standard deviation (``$\sigma_3$'') of FWHM and $\alpha$ for the 
three fits to a single object.

  Neither the FWHM nor the $\alpha$ values returned by ISHAPE show any 
systematic dependence on $\alpha_0$, and the mean fitted
FWHM and $\alpha$ are generally quite close to the input values.  For 
S/N$\approx$50, 
the \rms\ scatter of the FWHM and $\alpha$ measurements are typically 
0.15--0.20 
pixels and 0.10-0.15 (dimensionless units), respectively. The mean standard 
deviation ($\sigma_3$) of the three individual fits with different $\alpha_0$
is much smaller than the object-to-object \rms\ scatter, 
indicating that
the uncertainty due to a particular initial guess is smaller
than the random measurement error. For typical cluster-like objects
with S/N$\approx$50, these tests suggest that the FWHM and $\alpha$
parameter can be measured with an accuracy $\sim20$\% and
10--15\%, respectively on a single image. 


  Note, however, that the error on the \emph{effective radius} \reff\ 
derived from the FWHM and $\alpha$ can be very much larger, especially if 
$\alpha\approx1$. It is therefore important that any error selection be 
carried out on FWHM and $\alpha$, and not on the derived quantity \reff . 

  As a final caution, the tests carried out here do not take into account 
any uncertainty on the PSF itself. In practice, this can be important even 
for HST images, since the so-called ``diffusion'' kernel makes an important 
contribution to the scattering of light between neighbouring pixels. While 
this effect is included in the modelling done by ISHAPE, the diffusion 
kernel has so far only been properly characterised for the F555W filter. 
It is probably appropriate also for the F547M and F606W filters, but for 
bluer or redder passbands the true diffusion kernel could be 
significantly different.

\section{Contamination in the ground-based sample}
\label{sec:gb_cmp}

\begin{table*}
  \caption{Clusters from ground-based survey reidentified in HST images.
   Photometry is from ground-based data. FWHM is the 
   full-width-at-half-maximum in pc derived from EFF $\alpha=1.5$ fits.  The 
   first 10 rows of the table are reproduced here; the full table (330 rows) 
   is only available in electronic form.}
\label{tab:gb}
\begin{tabular}{lrrrrrrccr} \hline
ID & $V$ & $U-B$ & $B-V$ & $V-I$ & $N$ & FWHM & Type & Fit ok? & Comments \\ \hline
n1156-296 & $19.92\pm0.02$ & $-0.19\pm0.03$ & $ 0.12\pm0.03$ & $ 0.18\pm0.03$ & 1 & $ 0.20\pm0.00$ & 1 & Y & - \\
n1156-310 & $20.66\pm0.05$ & $-0.10\pm0.09$ & $ 0.18\pm0.07$ & $ 0.57\pm0.07$ & 2 & $ 6.58\pm2.38$ & 1 & N & - \\
n1156-331 & $19.85\pm0.04$ & $-0.67\pm0.05$ & $-0.03\pm0.05$ & $ 0.20\pm0.07$ & 1 & $ 0.08\pm0.00$ & 1 & Y & - \\
n1156-348 & $18.43\pm0.01$ & $-0.79\pm0.02$ & $ 0.12\pm0.02$ & $ 0.61\pm0.02$ & 1 & $ 4.92\pm0.00$ & 2 & N & a,b \\
n1156-356 & $20.27\pm0.02$ & $-0.06\pm0.04$ & $ 0.31\pm0.03$ & $ 0.55\pm0.02$ & 1 & $ 8.64\pm0.00$ & 2 & N & k \\
n1156-361 & $18.69\pm0.01$ & $-0.88\pm0.02$ & $-0.03\pm0.02$ & $ 0.45\pm0.02$ & 1 & $ 1.24\pm0.00$ & 1 & Y & - \\
n1156-403 & $20.47\pm0.07$ & $-0.31\pm0.13$ & $ 0.24\pm0.11$ & $ 0.50\pm0.09$ & 2 & $ 7.20\pm4.74$ & 3 & N & - \\
n1156-433 & $20.45\pm0.07$ & $-0.24\pm0.11$ & $ 0.28\pm0.10$ & $ 0.68\pm0.09$ & 2 & $ 0.30\pm0.24$ & 2 & N & c \\
n1156-441 & $20.01\pm0.04$ & $ 0.27\pm0.10$ & $ 0.23\pm0.06$ & $ 0.42\pm0.07$ & 2 & $ 4.98\pm1.46$ & 1 & Y & - \\
n1156-442 & $19.92\pm0.08$ & $-0.54\pm0.09$ & $-0.09\pm0.10$ & $ 0.17\pm0.12$ & 2 & $ 3.86\pm0.08$ & 1 & Y & - \\
\hline
\end{tabular}
\end{table*}

\begin{table}
\caption{Contamination of ground-based sample for individual galaxies.
  N$_{\rm tot}$ is the total number of cluster candidates identified in the
  ground-based surveys, N$_{\rm HST}$ is the subset of those candidates covered
  by HST images and N$_{<0.5 {\rm pc}}$ is the number of unresolved
  sources. The adopted distance moduli are also listed (see Paper~I for 
  references).}
\label{tab:contam}
\begin{tabular}{lllllll} \hline 
  Galaxy   & $m-$M & N$_{\rm tot}$ & N$_{\rm HST}$ & N$_{<0.5 {\rm pc}}$ & 
    \multicolumn{2}{c}{L$_{\rm HST}$/L$_{>0.5 {\rm pc}}$} \\ 
           &               &               &          &          & $V$ & $U$ \\
  \hline
NGC  628 & 29.6 &  38 &  22 &   3 &  0.84 &  0.82 \\
NGC 1156 & 29.5 &  22 &  21 &   5 &  0.85 &  0.86 \\
NGC 1313 & 28.2 &  45 &  20 &   2 &  0.81 &  0.68 \\
NGC 2835 & 28.9 &   9 &   8 &   6 &  0.20 &  0.17 \\
NGC 2997 & 29.9 &  34 &   8 &   1 &  0.96 &  0.98 \\
NGC 3184 & 29.5 &  13 &   9 &   3 &  0.79 &  0.83 \\
NGC 3621 & 29.1 &  45 &  29 &   6 &  0.79 &  0.75 \\
NGC 4258 & 29.47 &  44 &  26 &   2 &  0.91 &  0.91 \\
NGC 5055 & 29.2 &  24 &  13 &   3 &  0.77 &  0.76 \\
NGC 5194 & 29.6 &  69 &  33 &   2 &  0.99 &  0.99 \\
NGC 5204 & 28.4 &   7 &   6 &   5 &  0.23 &  0.26 \\
NGC 5236 & 27.9 & 149 &  29 &   9 &  0.79 &  0.82 \\
NGC 5585 & 29.2 &   9 &   7 &   5 &  0.49 &  0.60 \\
NGC 6744 & 28.5 &  18 &   4 &   0 &  1.00 &  1.00 \\
NGC 6946 & 28.9 & 107 &  78 &  11 &  0.88 &  0.90 \\
NGC 7424 & 30.5 &  10 &   6 &   0 &  1.00 &  1.00 \\
NGC 7793 & 27.6 &  20 &  11 &   6 &  0.46 &  0.53 \\
\hline
\end{tabular}
\end{table}

  One of the motivations for this study was to test the reliability  of
the ground-based cluster identifications and quantify how much any
mis-identifications might affect the specific luminosities for the
cluster systems derived in Paper III. To this aim,
  Table~\ref{tab:gb} lists photometry and object classifications
for all cluster candidates originally detected in the ground-based surveys,
which are also included in the HST datasets. Of the 330 objects listed
in Table~\ref{tab:gb}, 192 are also included in Table~\ref{tab:all}.
The remaining 138 objects were excluded from the second round of
fits (and Table~\ref{tab:all}) because of too low S/N for the
variable-$\alpha$ fits.
However, size information is listed for all objects in Table~\ref{tab:gb} 
based on the first round of profile fits (with $\alpha=1.5$).
Because only one fit was done for each exposure,
meaningful error estimates are only available for cluster candidates with 2 
or more detections. The mean and median error on the cluster FWHM values
for objects with multiple measurements are 0.5 pc and 0.26 pc, respectively, 
or 0.1--0.2 pixels (for typical galaxy distances of 5 Mpc). 
This is consistent with the estimates of the accuracy of ISHAPE fits 
in Sec.~\ref{sec:ishape_tests} (see also Paper II).

\begin{figure}
\centering
\includegraphics[width=85mm]{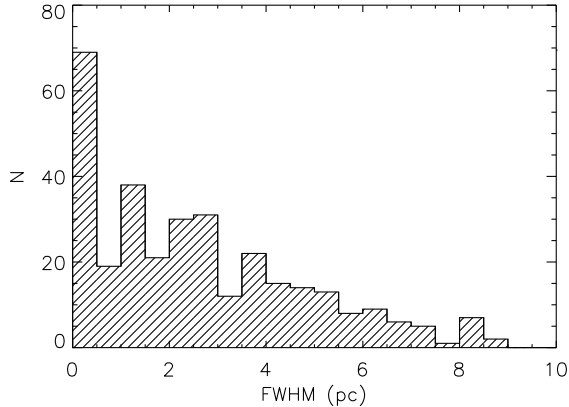}
\caption{Size distribution for cluster candidates identified in ground-based
  surveys.}
\label{fig:szhist_gid}
\end{figure}

\begin{figure}
\centering
\includegraphics[width=85mm]{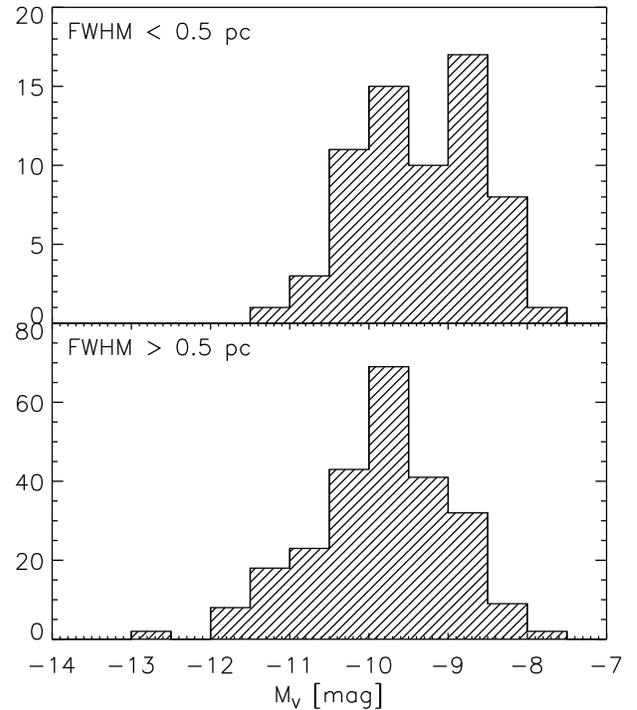}
\caption{Magnitude distributions for ground-selected cluster
  candidates with FWHM$<$0.5 pc (top) and FWHM$>$0.5 pc (bottom). 
  Sizes are from EFF fits with fixed $\alpha=1.5$.}
\label{fig:maghist_gb}
\end{figure}

\begin{figure}
\begin{minipage}{20mm}
n1313-363 \\
\includegraphics[width=20mm]{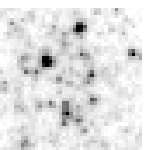}
\end{minipage}
\begin{minipage}{20mm}
n3621-1497 \\
\includegraphics[width=20mm]{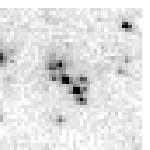}
\end{minipage}
\begin{minipage}{20mm}
n6946-1490 \\
\includegraphics[width=20mm]{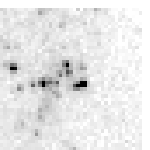}
\end{minipage}
\begin{minipage}{20mm}
n6946-3859 
\includegraphics[width=20mm]{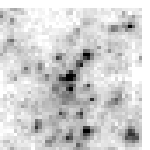}
\end{minipage}
\caption{Examples of objects identified as clusters from the ground,
  but which are unresolved according to the ISHAPE fits}
\label{fig:misclass}
\end{figure}

  The distribution of FWHM values for objects listed in Table~\ref{tab:gb} 
is shown in Fig.~\ref{fig:szhist_gid}.  Only 6 objects were classified as 
'likely/certain non-clusters' during the visual inspection, but
69 objects (or about 21\% of the sample) have FWHM$<$0.5 pc and are 
thus essentially unresolved even on the HST images. 
Unresolved objects are not necessarily individual, isolated stars, but 
can also be loose groupings of stars (OB associations etc.) where 
ISHAPE simply picks one star and fits it.  
This is the case for about 25 out of the 69 unresolved objects, 
or about 1/3. A few examples are shown in Fig.~\ref{fig:misclass}.
  The size distribution in Fig.~\ref{fig:szhist_gid} does not appear
strongly peaked at a particular value (except for the unresolved 
sources near FWHM$=$0), but spans a range from the resolution limit
up to 8--10 pc.  If objects with FWHM$<$0.5 pixels are excluded, the 
formal estimate of the mean FWHM is 3.8 pc. 
For an EFF model with 
$\alpha=1.5$, this corresponds to a half-light radius of 4.3 pc (or
4.0 pc if the profile is truncated at 50 pc), consistent 
with the typical sizes of young star clusters in the Milky Way and elsewhere.
The mean FWHM changes only slightly (to 3.9 pc) if objects with 
'Fit OK? = N' are excluded. The issue of cluster sizes will be discussed 
in more detail below for the full sample.

  Figure~\ref{fig:maghist_gb} shows the distributions of absolute $M_V$ 
magnitudes for unresolved and resolved objects. While the magnitude 
distribution for resolved objects (bottom panel) does extend to brighter 
magnitudes than for unresolved ones, there are several unresolved 
objects brighter than $M_V=-10$. Such bright objects are unlikely to be 
individual stars, but most of them have the 'c' comment set (companions 
within $1\farcs5$), indicating that the ground-based magnitudes are
likely contaminated by 
nearby objects. Another possibility is that some of them are very compact
star clusters.  At magnitudes fainter than $M_V=-9$,
Fig.~\ref{fig:maghist_gb} shows a
clear excess of unresolved objects
(26 out of 66, or 39\%, compared to 21\% for the whole sample),
many of which may indeed be individual stars. 


  Another way of estimating the contamination fraction is to use
the object types from the visual inspection. Of the 330 objects, 61 
are labelled as type `2' (uncertain) or `3' (very likely non-cluster).
Although classification as type `2' does not necessarily mean that the
object in question is a non-cluster, these numbers again suggest a
contamination rate of order 20\% or less. 

\begin{figure}
\centering
\includegraphics[width=85mm]{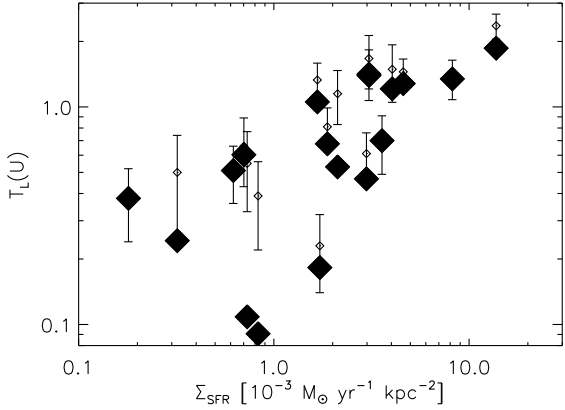}
\caption{Specific $U$-band luminosity ($T_L(U)$) vs.\ area-normalised
star formation rate (\ssfr ) for original ground-based
data and updated values (filled diamonds) based on contamination rates
estimated from HST data (Table~\ref{tab:contam}).}
\label{fig:tlu}
\end{figure}

  How does contamination of the cluster samples affect the
relations between specific luminosity ($T_L(U)$) and host galaxy 
properties derived in Paper III? If the contamination rate were the
same in different galaxies, all specific luminosities would just be 
scaled down by a constant factor, leaving any relations involving 
$T_L(U)$ intact. However, it is possible that the ratio of clusters
to potential contaminants varies from galaxy to galaxy.  
Table~\ref{tab:contam} 
lists the total number of cluster candidates identified from the ground 
(N$_{\rm tot}$) in each galaxy, followed by the number of ground-identified 
candidates contained within the WFPC2 images covering that galaxy 
(N$_{\rm HST}$), and the number of objects among these that are unresolved
(N$_{<0.5 {\rm pc}}$).
Based on the ground-based photometry, the
$V$- and $U$-band luminosities in resolved objects, relative to all
objects in the HST images, are listed in the last two columns. With a couple 
of exceptions, exclusion of unresolved objects affects the luminosities 
of the cluster systems only weakly. The two most notable 
exceptions are NGC~2835 and NGC~5204. In Paper I, we noted that all 7 
clusters in NGC~5204 are quite red and resulted in an unusually high
specific frequency (number of clusters relative to the host galaxy
luminosity). Now, it appears that most of these objects may in fact be
foreground stars.

  An updated version of the $T_L(U)$ vs.\ \ssfr\ plot
from Paper III is shown in Figure~\ref{fig:tlu}. The original data 
are shown with error bars, while new updated $T_L(U)$ values,
corrected according to Table~\ref{tab:contam}, are shown with filled
diamonds. The basic trend for $T_L(U)$ to increase as a function
of \ssfr\ is clearly preserved.
The scatter increases somewhat after the exclusion of unresolved objects.
However, with only a couple of clusters left in galaxies 
like NGC~2835 and NGC~5204 after exclusion of unresolved objects, 
$T_L(U)$ as defined in Paper III is probably no longer a good approximation 
to the true relative luminosity of the cluster system. To obtain a more 
useful number, one would likely have to probe to fainter magnitudes and 
thereby sample the cluster population more completely.

\section{The full sample}
\label{sec:full_sample}

\subsection{Cluster ages}

  The broad-band colours of simple stellar populations (such as star 
clusters) are functions of both age and metallicity, with additional
complications arising from unknown reddenings and stochastic
effects due to the finite number of stars in a cluster (e.g.\
Girardi et al.\ \cite{gir95}). However, for
clusters younger than $\sim10^9$ years it is still possible
to obtain reasonably accurate photometric age estimates, especially if 
$U$-band data are included. For such clusters, metallicity effects
are weak, except for a brief period around $10^7$ years when the
integrated light is dominated by red supergiants.  
Here, cluster ages were obtained by fitting Bruzual \& Charlot 
(2000; priv.\ comm.) model 
colours to the observed $UBVI$ cluster colours.  The SSP model fits were done 
by minimizing the \rms\ deviation between model- and observed colours (weighted by their errors) as a function of age and reddening.  In order to reduce the 
uncertainty on the age determinations, age estimates were made only for 
clusters with $\sigma_{B-V} < 0.2$ mag, $\sigma_{V-I} <0.2$ mag and 
$\sigma_{U-B} < 0.3$ mag.

  The ages of individual clusters derived from broad-band colours should 
only be regarded as approximate. The ground-based apertures may be 
contaminated by objects other than the cluster candidate itself, and model 
uncertainties also make the absolute ages uncertain. However, the relative age 
ranking of clusters should still be reasonably reliable. Line emission can 
also affect the broad-band colours of very young objects, and must be taken into
account if accurate age estimates for objects younger than $\simeq10^7$
years are required (e.g.\ Anders \& Fritze-v.\ Alvensleben \cite{af03}).

\subsection{Cluster sizes}
\label{sec:cluster_sizes}

\begin{figure}
\centering
\includegraphics[width=85mm]{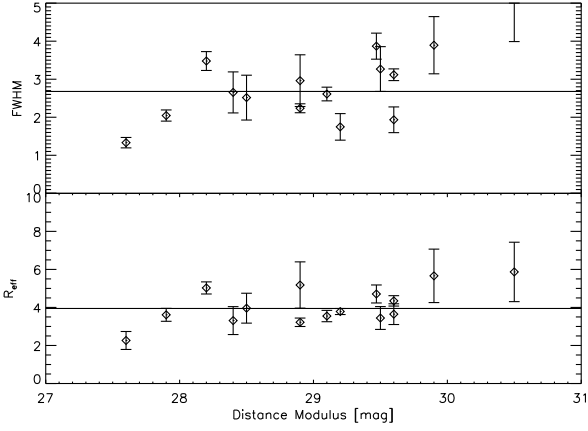}
\caption{Mean FWHM (top) and effective radii (bottom) for
  stellar clusters versus parent galaxy distance modulus.}
\label{fig:sz_dist}
\end{figure}

\begin{figure}
\centering
\includegraphics[width=85mm]{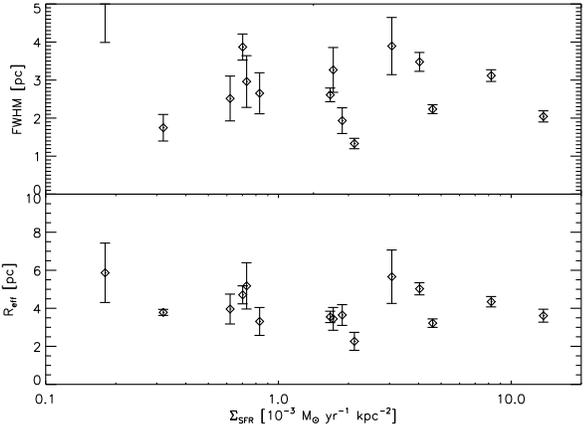}
\caption{Mean FWHM (top) and effective radii (bottom) for
  stellar clusters versus parent galaxy \ssfr\ (derived from IRAS 
  far-infrared luminosities, as described in Paper III).}
\label{fig:sz_sfr}
\end{figure}

\begin{table}
\caption{Mean FWHM and effective radii for clusters in galaxies.}
\label{tab:sztab}
\begin{tabular}{llccc} \hline 
  Galaxy   & N & $\langle$FWHM$\rangle$ & $\langle R_{\rm eff}\rangle$ ($\alpha>1$) & $\langle R_{\rm eff}\rangle$ (all) \\
           &   &     pc                 &     pc  & pc \\ \hline
NGC 628 & 30 & $1.93\pm0.34$ & $3.65\pm0.55$ & $4.82\pm0.71$ \\
NGC 1313 & 67 & $3.48\pm0.25$ & $5.03\pm0.32$ & $5.83\pm0.44$ \\
NGC 2835 & 8 & $2.96\pm0.68$ & $5.18\pm1.22$ & $5.18\pm1.22$ \\
NGC 2997 & 11 & $3.89\pm0.75$ & $5.66\pm1.41$ & $8.73\pm1.75$ \\
NGC 3184 & 15 & $3.27\pm0.59$ & $3.44\pm0.60$ & $7.08\pm1.71$ \\
NGC 3621 & 49 & $2.61\pm0.18$ & $3.55\pm0.30$ & $4.37\pm0.48$ \\
NGC 4258 & 44 & $3.87\pm0.34$ & $4.71\pm0.47$ & $6.45\pm0.66$ \\
NGC 5194 & 126 & $3.12\pm0.15$ & $4.34\pm0.27$ & $5.48\pm0.36$ \\
NGC 5204 & 16 & $2.65\pm0.54$ & $3.31\pm0.73$ & $4.33\pm1.22$ \\
NGC 5236 & 80 & $2.04\pm0.15$ & $3.61\pm0.34$ & $4.84\pm0.47$ \\
NGC 5585 & 6 & $1.75\pm0.35$ & $3.79\pm0.16$ & $3.79\pm0.16$ \\
NGC 6744 & 10 & $2.52\pm0.59$ & $3.96\pm0.79$ & $3.96\pm0.79$ \\
NGC 6946 & 131 & $2.24\pm0.12$ & $3.22\pm0.22$ & $4.57\pm0.35$ \\
NGC 7424 & 12 & $5.24\pm1.25$ & $5.87\pm1.56$ & $5.87\pm1.56$ \\
NGC 7793 & 55 & $1.33\pm0.14$ & $2.26\pm0.47$ & $4.68\pm1.01$ \\
All      & 661 & $2.68\pm0.07$ & $3.94\pm0.12$ & $5.24\pm0.17$ \\
\hline
\end{tabular}
\end{table}

  Table~\ref{tab:sztab} lists the mean FWHM and \reff\ for clusters in 
each galaxy.  Only clusters classified as 'Type 1' and with 
'Fit OK = YES' were included.  The number of clusters in each galaxy 
satisfying these criteria is given in the second column. Because of the
poorly defined \reff\ for clusters with $\alpha<1$, such clusters
were excluded before computing the average $\reff$ values in
Table~\ref{tab:sztab}.
Objects with very steep 
envelope slopes ($\alpha>5$) represent clear outliers (see 
Section~\ref{sec:struct_env} below) and were also excluded from further 
analysis.  The mean values given 
in the table are unweighted. This is because there is a correlation between 
the cluster sizes and their associated errors (the \emph{relative} errors 
remain roughly constant), which would lead to a strong bias in the mean if 
the size measurements were error-weighted.  

  The mean effective radius is 
$\langle \reff \rangle$ = $\meanreff \pm \emeanreff$ pc,
perhaps slightly larger compared to those for the ground-selected sample 
(Sec~\ref{sec:gb_cmp}) and other young and old star clusters. This may be due 
to the fact that the clusters are assumed to follow a single power-law out to 
a total radius of 50 pc, while in reality the behaviour at large radii is 
poorly constrained.  If the cluster profiles decline more rapidly at large 
radii the effective radii would decrease, especially for objects with 
$\alpha$-values close to 1.  Also, the size cut imposed in order to 
exclude point sources may introduce a bias against the most compact clusters.
 In the last column of Table~\ref{tab:sztab}, the $\alpha>1$
 requirement is abandoned.  Clearly, this leads to an increase in the
 mean effective radii, but it is stressed that the \reff\ values
 for these objects are very uncertain and depend strongly on the
 adopted outer radii. 

  The mean FWHM and \reff\ from Table~\ref{tab:sztab} are plotted versus 
parent galaxy distance modulus in Fig.~\ref{fig:sz_dist}.
The sizes do show some correlation with parent galaxy distance, 
possibly due to a less than perfect correction for the PSF. It 
is also possible that a larger fraction of the objects detected
in more nearby galaxies are individual stars, rather than clusters, which
made it into the list of cluster candidates despite the size cut.  
Contributing to this effect, the number of individual stars bright enough 
to be detected in the ground-based photometry would increase at 
smaller distances.  
Furthermore, the fixed size cut at FWHM$=$0.2 pixels corresponds to 
a different physical cluster size in different galaxies, ranging from 0.3 pc 
(core radius $\sim0.15$ pc) in NGC~7793 to 1.2 pc (core radius $\sim0.6$ pc) in 
NGC~7424.

  Fig.~\ref{fig:sz_sfr} shows the mean cluster sizes as a function of
the area-normalised host galaxy star formation rate, \ssfr\ (see Paper III). 
The scatter in the FWHM plot is somewhat larger than for the \reff\
values, 
but neither 
FWHM nor \reff\ shows any obvious correlation with \ssfr.  Below, the
data for all galaxies is combined and analysed collectively in order to
improve statistics, but in order to reduce possible systematic effects 
due to different distances, clusters in the closest (NGC~7793) and two most 
distant galaxies (NGC~2997 and NGC~7424) are excluded for the analysis of 
structural parameters. Thus, in summary the selection parameters applied to 
the cluster candidates in Table~\ref{tab:all} for the following analysis are:
\begin{itemize}
  \item Object type = 1 (certain / very likely cluster)
  \item Fit = OK
  \item When \reff\ is involved: require $1<\alpha<5$
  \item Photometric errors: $\sigma_{B-V} < 0.2$ mag, 
        $\sigma_{V-I} <0.2$ mag and $\sigma_{U-B} < 0.3$ mag
  \item Exclude NGC~2997, NGC~7424 and NGC~7793
\end{itemize}

\begin{figure}
\centering
\includegraphics[width=85mm]{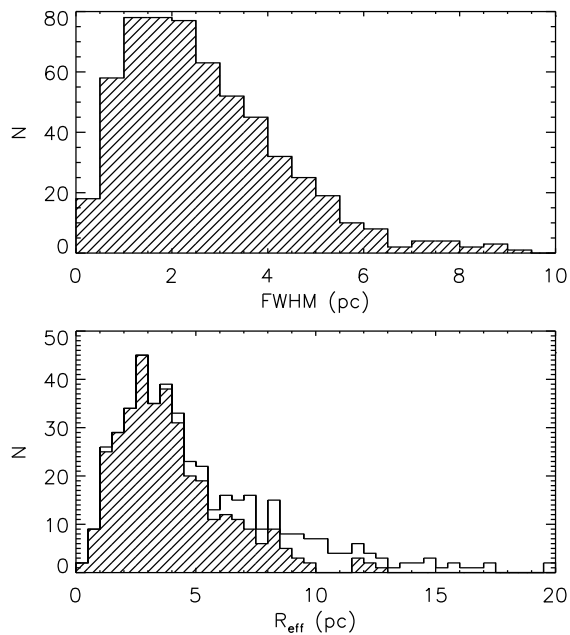}
\caption{Distributions of FWHM and effective radii for the combined
  cluster samples. In the lower panel, the hatched
  and outlined histograms are for clusters with $1<\alpha<5$ and
  $\alpha<5$, respectively.}
\label{fig:szhist}
\end{figure}

  Figure~\ref{fig:szhist} shows the distributions of FWHM and \reff\
values for the combined sample. The paucity of objects with FWHM$<1$ pc
is partly a selection effect, due to the size cut at FWHM=0.2 pixels, so
it is not clear that the distribution of FWHM even has a well-defined
peak. However, most clusters have FWHM$<$6--8 pc, similar to the range
spanned by the ground-selected clusters (Fig.~\ref{fig:szhist_gid}).
The \reff\ distribution is more complicated to interpret because the cut 
in FWHM does not translate to a unique \reff\ but depends on 
$\alpha$. For a typical $\alpha\sim1.3$ (see below), a FWHM of 1 pc 
corresponds to $\reff=1.6$ pc, but this increases to $\reff=3$ pc for
$\alpha=1.1$ and $\reff=8$ pc for $\alpha=0.9$. However, there is a
fairly rapid drop-off in the distribution for $\reff>5$ pc, and the vast
majority of the clusters have $\reff<10$ pc.  The mean FWHM and \reff\ 
values given in Table~\ref{tab:sztab} and shown in Figs.~\ref{fig:sz_dist} 
and \ref{fig:sz_sfr} should clearly be interpreted with caution.



\subsection{Envelope slope versus age}
\label{sec:struct_env}


\begin{figure}
\centering
\includegraphics[width=85mm]{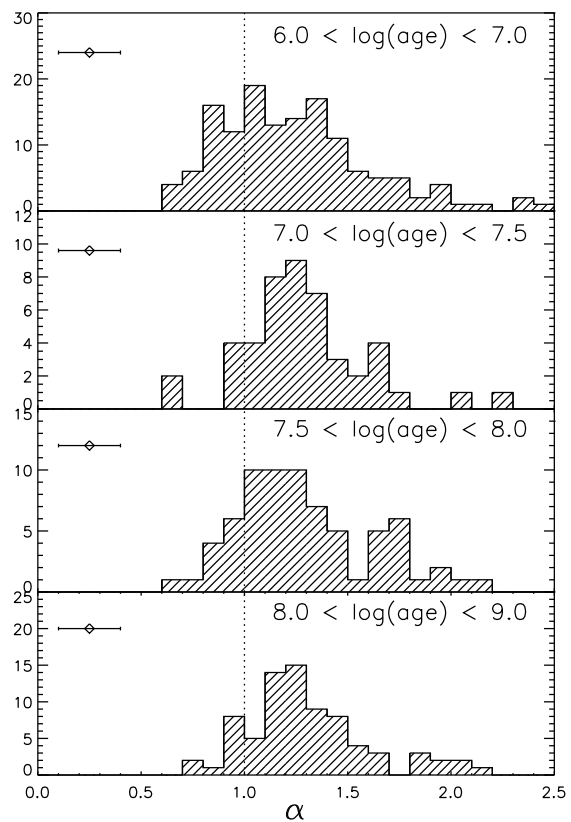}
\caption{Distribution of envelope slopes in four age bins}
\label{fig:hist_pow}
\end{figure}

\begin{figure}
\centering
\includegraphics[width=85mm]{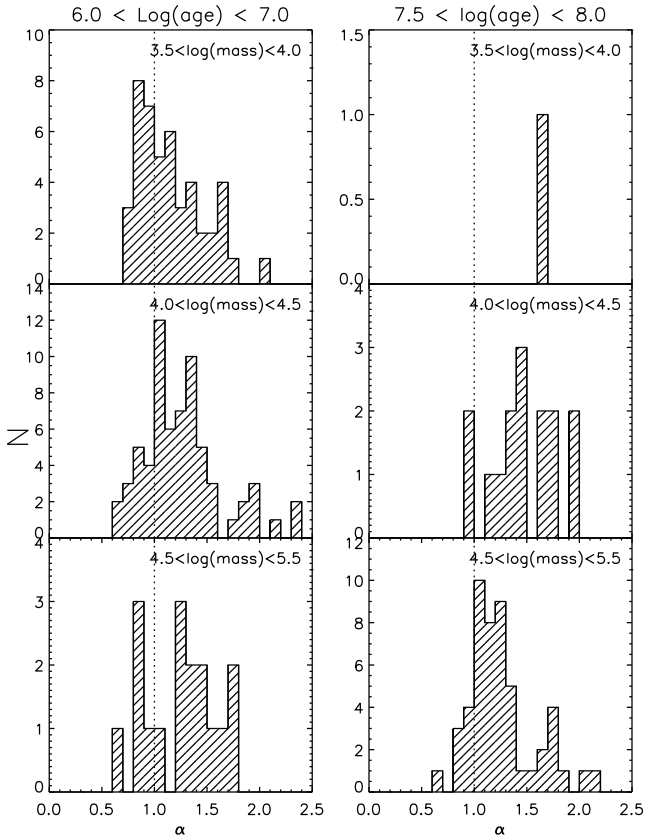}
\caption{Envelope slopes in different mass bins for two different ages}
\label{fig:pow_mass}
\end{figure}

  Figure~\ref{fig:hist_pow} shows the distribution of $\alpha$ values for 
clusters in four different age bins. For reference, typical errors of 0.15 
on $\alpha$ are indicated (cf.\ Sec~\ref{sec:ishape_tests}).  The distribution 
of $\alpha$ values is peaked around 1.2--1.3 in all four bins, but there may 
be a trend of mean $\alpha$ increasing with age,
at least in the sense that the youngest age bin appears to contain a
higher fraction of clusters with $\alpha<1$.
One important caveat in this comparison is that the mean mass is likely to 
increase from the youngest to the older age bins, due to the increase in 
M/L ratio with age.
Thus, in principle the difference between the $\alpha$-distributions in
Fig.~\ref{fig:hist_pow} might be due to the different mass ranges sampled
in each bin, rather than being an evolutionary effect.
To test whether this might be the case, Fig.~\ref{fig:pow_mass}
displays the $\alpha$-distributions in two age intervals,
but now also divided into different mass intervals. The number of
clusters in each panel is quite small, but the excess of clusters with
$\alpha<1$ seems to be present for log(age)$<$7.0 in all three mass bins.  
This suggests that the difference between the $\alpha$-distributions in 
Fig.~\ref{fig:hist_pow} is not just an effect of the different mass
intervals covered at different ages.

  Other authors have previously noted that very young clusters tend
to be surrounded by relatively extended envelopes with more light at
large radii than a King profile. Elson et al.\ (\cite{els87}) estimated that 
as much as 50\% of the mass in young LMC clusters may reside in unbound halos.
In the Milky Way, a large fraction of the youngest open clusters also have 
very large radii and may be unbound and in the process of dispersing away 
(Janes et al.\ \cite{jan88}).
Whitmore et al.\ (\cite{whit99}) showed radial profiles for three
clusters in the Antennae, illustrating a gradual transition from
extended envelopes with no well-defined outer radius (for the
highly luminous 'Knot S', only a few Myr old) to older clusters where
a tidal cut-off becomes apparent. A similar extended halo was observed
for a very luminous, 15 Myr old cluster in NGC~6946 (Larsen et al.\ 
\cite{laretal01}). The structure of very young clusters may hold important 
clues to the structure of the progenitor clouds out of which the clusters
formed, although it may prove challenging to disentangle this from the
effects of early dynamical evolution.

\subsection{Cluster sizes versus age and mass}

\begin{figure}
\centering
\includegraphics[width=85mm]{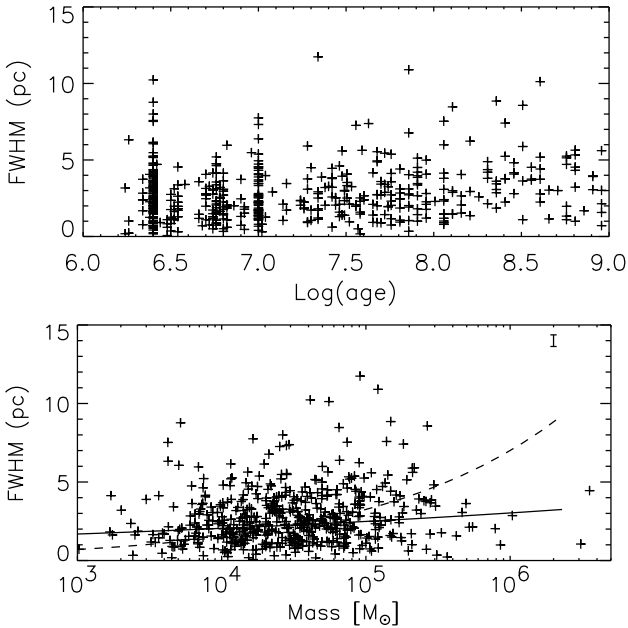}
\caption{Cluster FWHM versus age (top) and mass (bottom). The dashed curve
  in the lower panel represents a relation corresponding FWHM$\propto$M$^{1/3}$
  (constant density). The solid line is a least-squares power-law fit to
  FWHM vs.\ mass.}
\label{fig:fwhm_mass_age}
\end{figure}

\begin{figure}
\centering
\includegraphics[width=85mm]{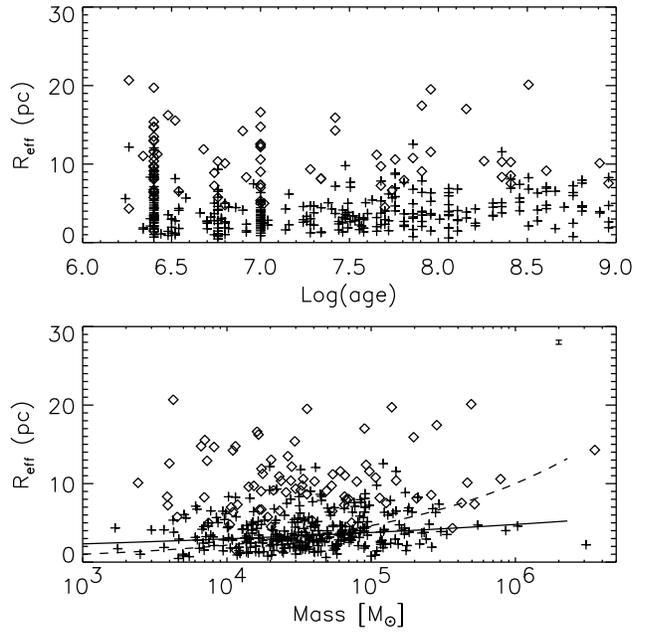}
\caption{Effective radius (\reff) versus age (top) and mass 
  (bottom panel). Clusters with $\alpha\leq1$ and $\alpha>1$ are
  shown with diamonds and plus markers, respectively.  The dashed curve
  in the lower panel represents a relation corresponding 
  $\reff \propto$M$^{1/3}$ (constant density). The solid line is a 
  least-squares power-law fit to $\reff$ vs.\ mass for clusters
  with $\alpha>1$.}
\label{fig:reff_mass_age}
\end{figure}

\begin{figure}
\centering
\includegraphics[width=85mm]{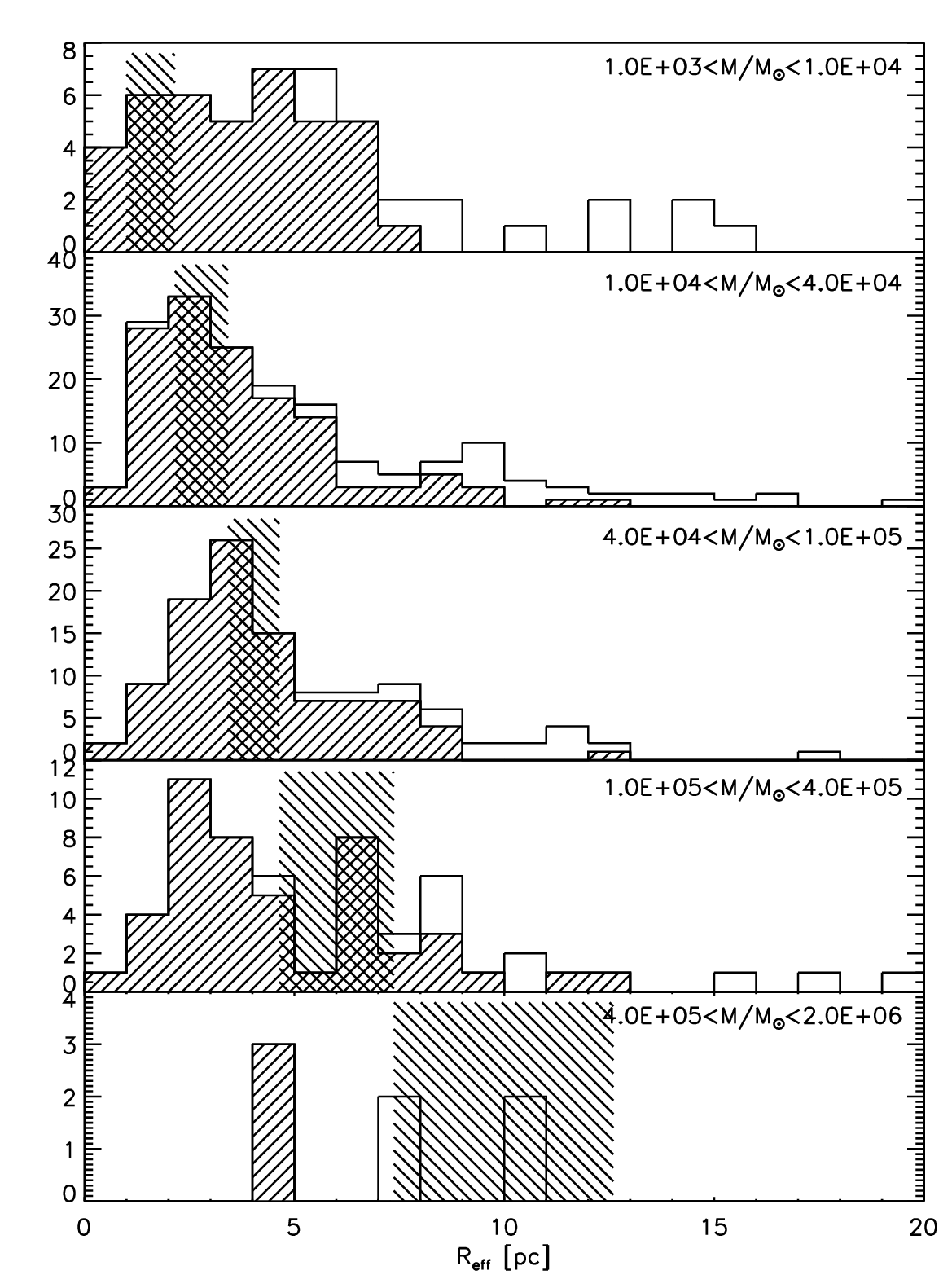}
\caption{Histograms of the \reff\ distributions in four mass ranges.
Filled histograms are for clusters with $\alpha>1$, while the
outlined histograms include clusters with $\alpha\leq1$. The shaded
areas indicate the range of sizes in each mass interval corresponding
to an $\reff \propto M^{1/3}$ relation.
}
\label{fig:reffhist}
\end{figure}

  Figures~\ref{fig:fwhm_mass_age} and \ref{fig:reff_mass_age} show FWHM 
and \reff\ versus age and mass for the combined cluster samples. Clusters 
older than 1 Gyr are excluded from 
the plots because of large uncertainties on the ages (and therefore also on 
the masses derived from integrated photometry).  The masses were estimated 
using M/L ratios from the Bruzual \& Charlot models, assuming a Salpeter IMF 
from 0.1 \msun -- 100 \msun.  Absolute masses are sensitive to the
shape and lower-mass cut-off of the IMF, 
and would be $\sim30$\% lower for a Kroupa (\cite{kroupa02}) IMF,
but this is not important for
the relative comparison attempted here.
In Fig.~\ref{fig:reff_mass_age}, clusters with $\alpha\leq1$ and 
$\alpha>1$ are shown with different symbols (diamonds and plus markers).
The dashed lines superimposed on the lower panels (size vs.\ mass) of each 
figure represent the size $\propto M^{1/3}$ relation corresponding to 
constant cluster density. 
Typical error bars are also shown in the top right corner in each of
the lower plots.


  Neither Fig.~\ref{fig:fwhm_mass_age} nor Fig.~\ref{fig:reff_mass_age}
shows any strong evidence for a general correlation between 
cluster size and age.  Since FWHM is roughly equivalent to 
$2\times r_c$, it is interesting to note that Mackey \& Gilmore 
(\cite{mg03}) found a strong correlation between core radius and age for 
clusters in the LMC (confirming earlier results by 
Elson et al.\ \cite{els89}). For ages $< 10^8$ years, there are essentially 
no clusters in the LMC with $r_c > 2.5$ pc, while clusters with ages 
$\sim10^9$ years show a full range of core radii from less than 1 to 8 pc.  
While it 
cannot be ruled out that the most extended clusters are missing from the
sample, Fig.~\ref{fig:fwhm_mass_age} shows no strong $r_c$-age relation 
similar to that in the LMC.  

Likewise, the size-mass plots show no strong correlations. The solid 
line in the bottom panel of Fig.~\ref{fig:fwhm_mass_age} is a least-squares 
power-law fit of the form
\begin{equation}
  {\rm FWHM} \, = \, A \, (M/\msun)^B
  \label{eq:fit_rc_m}
\end{equation}
  where $M$ is the mass and FWHM is measured in pc.  The best-fitting 
parameter values are $A=\constfwhm\pm\econstfwhm$ pc and 
$B=\slopefwhm\pm\eslopefwhm$.
  The fit formally indicates a $\sim3\sigma$ correlation between FWHM and 
mass, but with a large scatter.
The errors on the coefficients $A$ and $B$ are the formal errors from
the least-squares fit, but a jack-knife test yields nearly identical
error estimates.
The size cut may introduce a bias against more compact, low-mass objects, 
but the fit does not change substantially even if restricted to objects more 
massive than $10^4\msun$.  While no strong constraints can 
be put on a FWHM vs.\ mass relation based on Fig.~\ref{fig:fwhm_mass_age}, 
the formal relation derived from the least-squares fit is shallower than 
the $M^{1/3}$ relation corresponding to constant core density (indicated by 
the dashed line).
A similar fit is carried out for the
\reff\ vs.\ $M$ data in Fig.~\ref{fig:reff_mass_age}:
\begin{equation}
  \reff \, = \, A \, (M/\msun)^B,
  \label{eq:fit_re_m}
\end{equation}
  but only for clusters with $1<\alpha<5$.  Here, the best-fitting parameters 
are $A=\constreff\pm\econstreff$ pc and $B=\slopereff\pm\eslopereff$, quite 
similar to the slope of the FWHM vs.\ mass relation in 
Fig.~\ref{fig:fwhm_mass_age} and again only significant at about the $3\sigma$
level. Formally, the slope is slightly steeper than (but compatible with)
the $r\propto L^{0.07}$ relation found by Zepf et al.\ (\cite{zepf99}) for 
young clusters in NGC~3256.  As for the mean \reff\ values in 
Table~\ref{tab:sztab}, an 
important caveat is that the derived \reff\ are sensitive to the somewhat 
arbitrary truncation of the luminosity profiles at 50 pc. The slope of the 
\reff\ vs.\ $M$ relation remains stable at least for
cut-offs between 30 and 100 pc, 
even if the clusters with $\alpha<1$ are included in the fit
(but the normalisation changes somewhat).
As in the fit to FWHM vs.\ mass (and Table~\ref{tab:sztab}), the 
individual datapoints are \emph{not} weighted by their errors. 

  Using the HST photometry in F547M, F555W or F606W in Table~\ref{tab:hstphot} 
instead of the ground-based data to estimate the $V$-band luminosities 
of the clusters has no significant effect on the FWHM vs.\ $M$ and \reff\ vs.\
$M$ relations. For \reff\ vs.\ $M$, the slope $B$ decreases by only 0.003, 
while for FWHM vs.\ $M$ it decreases by 0.014, in both cases well within 
the formal uncertainties on the fits. Note, however, that the ages still
have to be estimated from the ground-based photometry.

The high degree of crowding in Fig.~\ref{fig:reff_mass_age} makes it 
difficult to visually assess to what extent the two curves agree with the 
data.  A somewhat different representation of the same data is shown in 
Fig.~\ref{fig:reffhist}, where the \reff\ distributions are shown for
five separate mass bins. The sample with $\alpha>1$ is shown with
shaded histograms, while the sample including clusters with $\alpha\leq1$
is shown with outlined histograms. In each panel, the range of sizes
corresponding to an $\reff\propto M^{1/3}$ scaling is also indicated
(using the same arbitrary normalisation as in Fig.~\ref{fig:reff_mass_age},
not taking any scatter into account). The paucity of objects in the
0--1 pc bin is due to the size cut, but even if one accounts for the
fact that the most clusters may have been systematically excluded, it
seems difficult to reconcile the overall \reff\ distributions in 
Fig.~\ref{fig:reffhist} with a constant-density relation.

  Although completeness effects are difficult to quantify, they would most 
likely tend to work against the detection of faint, extended clusters, thereby 
strengthening rather than weakening any existing size-mass trend. 
Thus, while the size-mass trends suggested by 
Figs.~\ref{fig:fwhm_mass_age} 
and \ref{fig:reff_mass_age} remain quantitatively uncertain, the current data 
seem to imply that any size-mass trend, if it exists, is substantially
shallower than for a constant-density relation. 

\subsection{Environment}
\label{sec:env}

\begin{table*}
\caption{Statistics on comment flags in different age intervals}
\label{tab:comment_flags}
\begin{tabular}{crrrrr} \hline
Age range                 & \multicolumn{2}{c}{Fit-OK = YES}
                          & \multicolumn{3}{c}{All} \\
                          & \multicolumn{2}{c}{Type = 1} & & \\
                        & a/c/d  & e/f/g/h & Type=2 & a/c/d  & e/f/g/h \\ \hline
6.0 $<$ log(age) $<$ 7.0  &   36\% &   2\% & 42\%   & 54\%   & 19\%    \\
7.0 $<$ log(age) $<$ 7.5  &   22\% &   0\% & 22\%   & 32\%   &  4\%    \\
7.5 $<$ log(age) $<$ 8.0  &   16\% &   0\% & 25\%   & 29\%   &  7\%    \\
8.0 $<$ log(age) $<$ 9.0  &   10\% &   0\% &  7\%   & 14\%   &  0\%    \\ \hline
\end{tabular}
\end{table*}

  While more subjective than the measurements of FWHM, $\alpha$ and
\reff , the comment flags in Table~\ref{tab:all} hold useful information 
about the surroundings of each cluster. 
It is reiterated that the comment flags are based on visual inspection
of a fairly heterogeneous dataset, and that the morphology of cluster
candidates can be quite wavelength-dependent. Note, however, that the
comment codes are based on visual inspection of all available images of
each cluster, and the majority of the clusters have imaging in a 
$B$-, $V$- or $I$-band equivalent filter.
  Table~\ref{tab:comment_flags} lists the fraction of cluster candidates with 
comment flags a/c/d (all of which are likely indicators of multiplicity) and 
e/f/g/h (more general crowding indicators). These statistics are given both 
for the best cluster candidates with structural parameters determinations 
(i.e.\ Fit-OK = YES and Type 1), as well as for all potential cluster 
candidates including those without reliable fits (Fit-OK = YES/NO and Type 
1/2). Data for all galaxies are included in 
Table~\ref{tab:comment_flags}, but the numbers remain unchanged within 
1--2\% even if NGC~2997, NGC~7793 and NGC~7424 are omitted as in 
the previous sections.

  Very few clusters with Fit-OK=YES have any of the /e/f/g/h flags set. 
This is no coincidence, because these flags indicate exactly those conditions 
which would make profile fits uncertain.  The table indicates a strong 
evolution in the environment as a function of age. Most of the cluster
candidates in 
crowded environments (flags e/f/g/h) are younger than $10^7$ years. The
tendency for the crowding to decrease with age is probably a consequence 
of fading and dispersion of the surrounding stellar population.  Assuming 
typical velocity dispersions of a few km/s within (unbound) star forming 
regions, the expansion will amount to a few tens of pc in $10^7$ years.

  Of the objects in the youngest age bin, 42\% were classified as 
``uncertain'' (type 2). This underscores the fundamental difficulty 
of identifying the youngest clusters. The problem of defining 
an appropriate selection criterion for \emph{bona-fide} clusters is far 
from trivial. Sometimes the main problem is simply that an object is only
barely resolved.  In such cases, better angular resolution would
help confirm or rule out the cluster nature. For objects with complex
morphology it can be difficult to determine whether or not an
object is a true star cluster, even if resolution would otherwise not
be a problem.  Examples can be seen in Fig.~\ref{fig:uncertain} and in 
panels (e), (f) and (g) of Fig.~\ref{fig:comm_ill}.  In these cases, it 
is difficult to determine whether a well-defined stellar cluster is present. 
For low-mass clusters of low age, an additional problem is that the 
integrated light can be dominated by a few luminous stars, making it
difficult to distinguish such objects from random superpositions of
individual stars along the line-of-sight. 

As pointed out in the introduction, most stars probably form in clusters,
but only a small fraction of young embedded clusters survive as bound
objects. It may also happen that only a fraction of the stars in a
cluster remain bound, while the rest disperse away (Kroupa \cite{kroupa01}). 
Thus, a few Myr-old concentration of stars may be a bound star cluster, a 
bound core surrounded by an expanding envelope, or an entirely unbound 
association which will soon disperse away completely. Other star formation 
may also be taking place nearby, perhaps triggered by the young cluster. So 
it is not surprising that a large fraction of the youngest objects have 
a messy morphology. 

  The age distribution of double or multiple objects does not appear
to be as strongly peaked at young ages, with some objects flagged 
'a/c/d' even in the oldest age bin. It is possible that at least a 
fraction of these objects are genuine double clusters, similar to those 
found in the Large Magellanic Cloud.  The LMC binary clusters tend to be 
predominantly young, though a few pairs as old as several times $10^8$ 
years exist (Dieball et al.\ \cite{dmg02}). This seems to be consistent with 
the decreasing fraction of multiple objects at high ages in 
Table~\ref{tab:comment_flags}.  However, because the main source of
photometry in this paper is ground-based imaging, no information is available 
about possible colour/age differences between the components in such pairs. 
Multi-colour HST imaging, especially including $U$-band data,
would allow a more thorough investigation of double 
clusters and make a comparison with the LMC sample possible.

\section{Summary and conclusions}

  Using a combination of HST/WFPC2 imaging and ground-based $UBVI$ photometry, 
a catalogue of cluster candidates in 18 nearby spiral galaxies has been 
compiled. Only objects with a S/N$>$50 on the HST images (within an $r=5$ 
pixels aperture) were included, allowing for a detailed analysis of the 
structure of individual clusters.  Analytic profile fits of the form $P(r) 
\propto (1+(r/r_c)^2)^{-\alpha}$ were carried out, including a proper 
modelling of the HST/WFPC2 PSF, and allowing both the core radius $r_c$ and
envelope slope parameter $\alpha$ to vary.  Each cluster candidate has
also been visually inspected and comment flags relating to crowding and
multiplicity are given. These comment flags, combined with 
the photometric data and structural parameters, may be helpful when using 
the list of cluster candidates to select targets e.g.\ for spectroscopic 
studies.

The HST imaging indicates a mean contamination rate of 20\% or less 
for the ground-based cluster surveys in Paper I--III,
although the contamination rates in some individual galaxies (most notably 
NGC~5204) are higher.  However, the relation between specific $U$-band 
luminosity of the cluster systems $T_L(U)$ and \ssfr\ (Paper III) remains 
valid after correction for contamination.  Because very few clusters
remain after the contamination correction in some galaxies, sampling the 
cluster populations to fainter magnitudes than the limits defined in
Papers I--III would probably reduce the scatter in the 
$T_L(U)$ vs.\ \ssfr\ relation. 

  The cluster catalogue has been used to investigate trends and relations 
between various cluster properties, although the analysis is complicated
by the fact that selection effects are difficult or impossible to control 
with a dataset as heterogeneous as the one used here.  
Most clusters have FWHM less than about 8 pc with a formal mean 
of about 2.7 pc, corresponding to a mean core radius 
$\langle r_c \rangle \approx$ 1.3 pc, but very compact clusters may be 
missing because of the size cut imposed in order to exclude point sources
(individual stars).
For the subset of the clusters which have $1<\alpha<5$ and therefore
reasonably reliable measurements of the effective radius \reff, the mean 
value is $\langle \reff \rangle = \meanreff \pm \emeanreff$ pc, but this
mean value may again be affected by a selection bias against the most
compact clusters. The effective radii are also
sensitive to the poorly constrained behaviour of the luminosity 
profiles at large radii. Here, the profiles are truncated at 50 pc.  
The structural parameters show little or no variation from 
galaxy to galaxy, especially when considering that the distances are not 
always known very accurately. In particular, the effective radii are 
uncorrelated with the host galaxy area-normalised star formation rate, and 
are also very similar to those of open and globular clusters in the 
Milky Way, globular clusters in early-type galaxies, and young clusters 
in merger galaxies and starbursts. It is quite remarkable that the 
sizes of stellar clusters are largely invariant with 
respect to the properties of the parent galaxy. Physical parameters such 
as gas density and -pressure probably play a major role in regulating the 
star formation rate (Kennicutt \cite{ken98}), but while these factors 
may affect the formation efficiency of bound clusters (Paper III) they do 
not appear to have a strong impact on the structure of the clusters 
themselves, once formed. Exceptions to this rule do exist, including
the ``faint fuzzy'' star clusters observed in some lenticular galaxies
(Brodie \& Larsen \cite{bl02}), and there is a general trend for the
sizes of globular clusters to increase as a function of galactocentric
distance (van den Bergh et al.\ \cite{van91})

While both the FWHM and \reff\ are found to correlate with cluster mass, 
least-squares power-law fits yield slopes that are substantially shallower 
than for a constant-density relation, implying an increase in cluster density 
as a function of mass.  
Qualitatively, these results are in agreement with previous data for
young star clusters as well as old globular clusters.
Quantitatively, the relations show a large scatter and remain uncertain. 
  Ashman \& Zepf (\cite{az01}) have argued that an increasing star formation
efficiency as a function of cluster mass may explain (at least partially)
the lack of a strong size-mass relation. 

  Many of the youngest clusters have extended, shallow outer envelopes.
This tendency seems to be a general one, noted 
previously for a few isolated cases in the Antennae, NGC~6946, and for LMC 
clusters.  The structure of these young objects may hold important clues
to the early dynamical evolution of clusters and the density distribution 
of the parent proto-cluster clouds. Older clusters gradually evolve towards 
King-type profiles with a finite tidal radius. 

  Finally, a strong correlation between cluster age and crowding is found, with
most of the strongly crowded clusters having ages $<10^7$ years. 
About 1/3-1/2 of these young objects are double or multiple sources, but 
the identification as bona-fide clusters is often uncertain even on
WFPC2 images in these fairly nearby galaxies. Future multi-colour
imaging with the \emph{Advanced Camera for Surveys} on HST should help 
resolve many of the difficulties encountered in this study.

\begin{acknowledgements}
  This work was partially supported by National Science Foundation
grant AST 02-06139 and by \emph{HST} grant AR-09523. I am grateful
to T.\ Richtler and S.\ M.\ Fall for several helpful comments, and to 
the referee, R.\ de Grijs, for a very detailed report which helped 
improve the paper.
\end{acknowledgements}

\end{document}